\documentclass[11pt,a4paper]{article}
\pdfoutput=1
\usepackage[utf8]{inputenc}
\usepackage[english]{babel}
\usepackage{amsmath, amssymb}
\usepackage{graphicx}
\usepackage{color} 
\usepackage{float}
\usepackage{xspace}
\usepackage{cancel}
\usepackage{tikz}
\usepackage{jcappub}
\usepackage[normalem]{ulem}

\definecolor{tabgreen}{HTML}{2ca02c}
\definecolor{tabblue}{HTML}{1f77b4}
\definecolor{tabred}{HTML}{d62728}

\newcommand{\gyr}{\,\text{Gyrs}}
\newcommand{\GeV}[0]{\text{GeV}}

\newcommand{\eV}[0]{\text{eV}}
\newcommand{\s}[0]{\text{s}}
\newcommand{\dx}{\,\text{d}} 

\newcommand{\lcdm}{\ensuremath{\Lambda}CDM\xspace}

\subheader{\small
	\vspace*{-1.28cm}
	\begin{flushright}
		TUM-HEP-1573/25
	\end{flushright}
}

\title{
Connecting cosmologically decaying dark matter to neutrino physics 
}

\author{Lea Fuß,}
\author{Mathias Garny,}
\author{Alejandro Ibarra}

\affiliation{Technical University of Munich,  
TUM School of Natural Sciences,
Department of Physics,\\
	James-Franck-Str.\ 1, 
	85748 Garching, Germany}

\emailAdd{lea.fuss@tum.de}
\emailAdd{mathias.garny@tum.de}
\emailAdd{ibarra@tum.de}

\abstract{Dark matter decays into invisible particles can leave an imprint in large-scale structure surveys due to a characteristic redshift-dependent suppression of the power spectrum. We present a model with two quasi-degenerate singlet fermions, $\chi_1$ and $\chi_2$, in which the heavier state decays as $\chi_2 \to \bar \chi_1 \nu \nu$ on cosmological time-scales, and that also accommodates non-zero neutrino masses. Remarkably, for parameters that yield the correct dark matter abundance via freeze-in and reproduce the observed neutrino masses, dark matter decay can produce detectable signals in forthcoming large-scale structure surveys, a diffuse anti-neutrino flux accessible to JUNO, and a gamma-ray line within the energy range probed by COSI. 
Both the cosmological lifetime of $\chi_2$ as well as the small (radiatively induced) mass splitting among $\chi_{1,2}$ are a natural consequence of the mechanism of neutrino mass generation within this model. This highlights the potential role of large-scale structure surveys in probing some classes of neutrino mass models.}

\begin{document}

\maketitle
\clearpage

\section{Introduction}
\label{sec:intro}

The standard paradigm of cold and stable dark matter (DM) has been remarkably successful in explaining a vast number of cosmological observations, most notably the characteristics of the Cosmic Microwave Background (CMB) radiation and the large-scale structure of our Universe. However, modifications of this paradigm are not precluded neither theoretically nor observationally. In fact, many simple models of particle dark matter introduce new particles and/or new interactions which may lead to unstable or warm dark matter components in our Universe.  

Current observations severely constrain the dark matter decay widths into Standard Model (SM) particles (see for example~\cite{Cirelli:2025rky,Slatyer:2017sev,Ibarra:2013cra}). On the other hand, decays into dark sector particles are much less constrained. However, these decays can lead to modifications in the cosmological background and perturbations which can be tested by CMB and large-scale structure surveys (see \emph{e.g.}~\cite{Audren:2014bca,Enqvist:2015ara,Berezhiani:2015yta,Poulin:2016nat,Enqvist:2019tsa,Bringmann:2018jpr,Pandey:2019plg,DES:2020mpv,Nygaard:2020sow,Holm:2022kkd,Alvi:2022aam,Nygaard:2023gel} for works on massless decay products and~\cite{Aoyama:2011ba,Wang:2012eka,Wang:2014ina,Blackadder:2014wpa,Aoyama:2014tga,Blackadder:2015uta,Vattis:2019efj,PhysRevD.103.043014,Haridasu:2020xaa,Abellan:2020pmw,FrancoAbellan:2020xnr,FrancoAbellan:2021sxk,Simon:2022ftd,Fuss:2022zyt,Bucko:2023eix,Montandon:2025xpd,Peter:2010sz,DES:2022doi,Lester:2025hqt} for massive ones).
Moreover, dark matter decays into dark sector particles with a cosmologically long lifetime has recently received some attention as a possible explanation of apparent inconsistencies in the standard cosmological model (\lcdm) like the Hubble tension~\cite{PhysRevD.103.043014,Haridasu:2020xaa,Davari:2022uwd,Nygaard:2020sow,Nygaard:2023gel,Zhou:2025ikl}, the $S_8$ tension~\cite{Nygaard:2020sow,FrancoAbellan:2021sxk,Davari:2022uwd}, small scale problems~\cite{Wang:2014ina}, as well as the preference for negative neutrino masses~\cite{Lynch:2025ine} (see also~\cite{Giare:2025ath}).

We consider a class of decaying cold dark matter (DCDM) models in which the decay products populate both the visible and the dark sector. In principle, this scenario allows to correlate cosmological signals with indirect detection signatures, and possibly with dark-matter production, since the same interactions that induce the decay act as a portal between the two sectors. To evade the stringent constraints from indirect detection, the three-body decay $\chi_2 \rightarrow \bar \chi_1 \nu \nu$ emerges as the simplest possibility~\cite{Fuss:2024dam}, where $\chi_2$ and $\chi_1$ are quasi-degenerate singlet fermions. The mass difference between these DM particles is partially converted into kinetic momentum for $\chi_1$, which can impact structure formation by suppressing clustering on small scales. Moreover, if the mass splitting is sufficiently small, the phase space available for the neutrinos to decay is correspondingly small, suppressing their momenta. While current experiments are largely insensitive to a diffuse flux of low-energy (anti-)neutrinos, upcoming instruments such as JUNO~\cite{ColomerMolla:2023ppf} may be capable of detecting a signal in the same parameter region that also produces cosmologically observable signatures in the large-scale structure.

In this work we construct a model where the eight-dimensional effective field theory (EFT) operator leading to  $\chi_2\rightarrow \bar \chi_1 \nu \nu$ is generated after electroweak symmetry breaking by integrating-out heavy degrees of freedom. We  investigate the role of the heavy particles in the production of dark matter in the early Universe and in the generation of neutrino masses, thus allowing to correlate three seemingly disconnected phenomena: neutrino masses, indirect detection signals and deviations in the matter power spectrum in the large-scale structure of the Universe. 

This paper is structured as follows: In Sec.~\ref{sec:MinimalDCDM} we briefly summarize the decaying dark matter model inducing $\chi_2\rightarrow \bar \chi_1 \nu \nu$ that was introduced in~\cite{Fuss:2024dam} by following an EFT approach. In Sec.~\ref{sec:Neutrino_masses} we present a renormalizable model which at low energies leads to the dark matter decay $\chi_2\rightarrow \bar \chi_1 \nu \nu$, as well as to non-zero neutrino masses. 
In Sec.~\ref{sec:benchmarks} we discuss the observational consequences of the model in cosmology, neutrino  and laboratory experiments, and showcase a number of benchmark models. In Sec.~\ref{sec:L_violation_decays} we discuss the implications of the lepton number violating decays predicted by the model, such as  $\chi_2\rightarrow \bar \chi_1 e^+ e^-$ or $\chi_2\rightarrow \bar \chi_1 \gamma$ and lastly, in Sec.~\ref{sec:conclusion}, we summarize our findings. We also include appendices~\ref{sec:app_symmetries}, \ref{sec:app_mass_splitting} and \ref{sec:app_L_violating_decays},  presenting technical details of the model building, the calculation of the mass splittings, and the rates for the lepton number violating decays, respectively.

\section{Minimal decaying dark matter}
\label{sec:MinimalDCDM}

The impact of decaying dark matter on the large-scale structure of the Universe can be parametrized by two parameters: the lifetime $\tau$, which is typically larger than the age of the Universe, and the mass splitting between the two massive DM particles (with masses $M$ and $m$, $M>m$), which can be described by the dimensionless parameter
\begin{equation}\label{eq:defepsilon}
  \epsilon = \frac{1}{2}\left(1-\frac{m^2}{M^2}\right)\,.
\end{equation}
In the degenerate limit, $\epsilon\ll 1$, the cosmological evolution of the homogeneous background is largely unaffected, whereas the velocity kick  $v_{k}/c\simeq \epsilon$ received by the massive daughter particle can sizably affect the density perturbations, producing a late-time suppression in the matter power spectrum on small scales. Cosmological limits on decaying dark matter scenarios have been derived from CMB data~\cite{Abellan:2020pmw,FrancoAbellan:2021sxk,Simon:2022ftd}, galaxy clustering~\cite{Simon:2022ftd}, the Lyman-$\alpha$ forest~\cite{Fuss:2022zyt}, weak lensing~\cite{Bucko:2023eix,Montandon:2025xpd} as well as Milky Way satellite abundances~\cite{Peter:2010sz,DES:2022doi} (see also~\cite{Nadler:2025yni}) which point to a region in parameter space of interest broadly in the ballpark of $\tau\sim {\mathcal O}(100)\,$Gyrs and $\epsilon\sim {\cal O}(10^{-2})$.

\begin{figure}[t]
    \centering
    \includegraphics[width=0.165\textwidth]{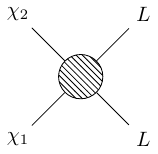}
    \hspace{1.5cm}
    \includegraphics[width=0.349 \textwidth]{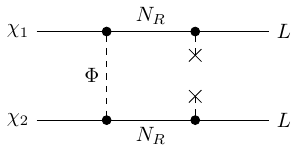}
    \caption{Left: Effective operator giving rise to the decay $\chi_2 \rightarrow \bar{\chi_1} \nu \nu$. Here $\chi_{1,2}$ are the DM states, and $L$ is the left-handed SM lepton doublet, being contracted with the SM Higgs field (that is not shown for brevity, see Eq.~\eqref{eq:Lagrangian}), resulting in a four-fermion interaction with the neutral neutrino component of $L$ after EWSB (see Eq.~\eqref{eq:Leff}).\\
    Right: Diagram generating the effective interaction within the UV complete model considered in this work. Here $N_R$ is a heavy sterile neutrino and $\phi$ a scalar mediator between the dark sector and the SM. The crosses symbolize insertions of the SM Higgs field $H$ (projecting out the $\nu$ component of $L$ when inserting the VEV after EWSB), and the corresponding HNL interaction is associated to neutrino mass generation.}
    \label{fig:UV_completion}
\end{figure}

In~\cite{Fuss:2024dam} a decaying DM scenario was introduced (see~\cite{Bell:2010fk,Bell:2010qt,Hamaguchi:2017ihw,Bae:2018mgq,Choi:2021uhy,Deshpande:2023gij,Obied:2023clp,Cheek:2022yof,Cheek:2024fyc,Cardenas:2024ojd} for related possibilities) which generates signals not only in the large-scale structure but also in neutrino experiments, while avoiding the very stringent constraints from indirect detection experiments. The minimal model contains two quasi-degenerate fermion singlets\footnote{Note that the DM states $\chi_{1,2}$ correspond to $N_{1,2}$ in the notation used in~\cite{Fuss:2024dam}. In this work, we instead reserve $N$ to denote right-handed neutrinos, following the common notation.}, $\chi_1$ and $\chi_2$, with masses $m$ and $M$ respectively ($M>m$), both with lepton number $+1$ and with opposite charges under a new global symmetry $U(1)_\chi$. With this assignment, $\chi_2$ can decay into $\bar \chi_1$ and two neutrinos through a unique gauge invariant dimension-eight operator
\begin{equation}
	{\mathcal L}_{\text{int}} = \frac{1}{\Lambda^4} \left(\bar{L}\tilde{H}P_R \chi_2\right) \left(\bar{L}\tilde{H}P_R \chi_1\right) + \text{h.c.}\,,
    \label{eq:Lagrangian}
\end{equation}
where $H$ is the standard Model Higgs doublet, $\Lambda$ is a high energy mass scale and $P_R$ is a projection operator. After electroweak symmetry breaking (EWSB), the following dimension-six operator is generated
\begin{equation}\label{eq:Leff}
    \mathcal{L}_\text{eff} = \frac{v^2}{2\Lambda^4} \, \bar{\nu} P_R \chi_2 \, \bar{\nu} P_R \chi_1\,,
\end{equation}
which in turn leads to the three body decay $\chi_2 \rightarrow \bar{\chi_1} \nu \nu$ with a rate of
\begin{equation}
\Gamma=\frac{v^4}{1280 \pi^3 \Lambda^8} (\epsilon M)^5\,,
\label{eq:Lambda}
\end{equation}
where $v$ means the typical Higgs vacuum expectation value (VEV). For $\Lambda\sim\mathcal{O}(50\,\text{TeV})$ and $M\sim\mathcal{O}(1\GeV)$, the predicted lifetime is in the ballpark of the values that can leave an imprint in the large-scale structure. Interestingly, these same parameters lead to a diffuse neutrino flux that could be detected by the JUNO experiment or other upcoming neutrino experiments. 

It is important to remark that the assignment of global charges forbids operators of the form $(\bar \chi_1 L)(\chi_2\bar L)$ or $(\bar \chi_1 \chi_2)(\bar L L)$, which would lead to the tree-level decay $\chi_2\rightarrow \bar \chi_1 e^+ e^-$ or the one-loop decay $\chi_1\rightarrow \bar \chi_1 \gamma$. The rate of these processes is constrained by experiments to be at least ten orders of magnitude longer than the age of the Universe~
\cite{DelaTorreLuque:2023olp,DelaTorreLuque:2023cef,Jin:2013nta}, which in turn makes the detection of a correlated dark matter decay signal in the large-scale structure of the Universe extremely challenging. The assignment of charges proposed in~\cite{Fuss:2024dam} thus guarantees that Eq.~\eqref{eq:Lagrangian} is the leading operator, and that correlated signals in the large-scale structure and in neutrino experiments are possible.

\section{UV completions and connection to neutrino mass generation}
\label{sec:Neutrino_masses}

In this section we construct two renormalizable models that generate the effective theory described in  Sec.~\ref{sec:MinimalDCDM} at low energies, and where the small mass splitting between the two fermionic singlets naturally arises as the consequence of a softly broken symmetry. To this end, we introduce additional heavy fermions and a complex scalar, all singlets under the Standard Model gauge group. The new heavy fermions are singlets under $U(1)_\chi$ and have lepton number $+1$, while the complex scalar is charged $+1$ under $U(1)_\chi$ and carries no lepton number.

The assignment of global charges allows Yukawa couplings between $\chi_{1,2}$ and the new fermions, which will generate the effective operator Eq.~\eqref{eq:Lagrangian}, as well as Yukawa couplings between the new fermions and the SM lepton doublets. Furthermore, we will allow lepton number breaking terms in the Lagrangian, which will play a pivotal role in the phenomenology of the model. Concretely, the lepton number breaking terms are responsible for generating a mass splitting between $\chi_1$ and $\chi_2$ via quantum effects. Furthermore, the same lepton number breaking terms generate non-zero neutrino masses, thus providing a natural connection between the effective theory constructed in~\cite{Fuss:2024dam} and neutrino masses. 

More specifically, we will consider two archetype scenarios of neutrino mass generation, namely the standard type I seesaw model~\cite{Mohapatra:1979ia,Minkowski:1977sc,Yanagida:1979as,Gell-Mann:1979vob}, and the inverted seesaw model~\cite{Wyler:1982dd,Mohapatra:1986aw,
Mohapatra:1986bd,Bernabeu:1987gr}. In this work we will consider for simplicity only one generation of fermions, although the extension to the realistic scenario with three generation of SM fermions is straightforward. Each model will be described separately in the next subsections.

\subsection{Type I seesaw scenario}
\label{sec:seesaw}

We add one right-handed neutrino, $N_R$, and one complex scalar $\phi$ to the particle content of~\cite{Fuss:2024dam}. The quantum numbers of the different particles under the global symmetries $U(1)_\chi$ and $U(1)_L$ are as follows:
\begin{center}
    \begin{tabular}{c|c|c|c|c}
         & $\chi_2$ & $\chi_1$ & $N_R$ & $\phi$\\
         \hline
        $U(1)_\chi$ & +1 & -1 & 0 & +1\\
         \hline
        $U(1)_L$ & +1 & +1 & +1 & 0\\
    \end{tabular}
\end{center}

The Lagrangian of the model is (apart from standard kinetic terms) given by an interaction term as well as two mass terms, one that conserves lepton number and another one that breaks the lepton number
\begin{align}
    \mathcal{L} &=  \mathcal{L}_\text{Yukawa} + \mathcal{L}^L_\text{mass} + \mathcal{L}^{\cancel L}_\text{mass}\,.
\end{align}
The interaction term compatible with the gauge and the global symmetries of the model is given by
\begin{equation}\label{eq:Yukawa}
    \mathcal{L}_\text{Yukawa} = y \bar{L} \tilde{H} N_R + y_1\bar N_R \chi_1 \phi + y_2\bar{\chi_2} N_R \phi + {\rm h.c.}\,,
\end{equation}
while the lepton number conserving mass term reads
\begin{equation}\label{eq:Lmass}
    \mathcal{L}^L_\text{mass} \supset - M (\bar \chi_1 \chi_1 + \bar \chi_2 \chi_2) -M_\phi^2\phi^\dagger\phi\,.
\end{equation}
Here we have assumed that both $\chi_{1,2}$ have the same Dirac mass $M$. This assumption could be justified if  $\chi_{1,2}$ are embedded into a multiplet with an extended dark sector symmetry group (see Appendix~\ref{sec:app_symmetries} for a possible realization based on $SU(2)$). 

Finally, the lepton number breaking part of the Lagrangian consists of a Majorana mass term for the right-handed neutrino
\begin{align}
    & \mathcal{L}^{\cancel L}_\text{mass} \supset -\frac{1}{2}M_R \bar N_R^c N_R\,.
\end{align}

After the breaking of the electroweak symmetry the mass eigenstates no longer coincide with the interaction eigenstates, due to the Yukawa coupling of $N_R$ to the SM lepton and Higgs doublets. The mass matrix of the singlet fermions of the model is given by
\begin{equation}
    \begin{pmatrix}
    \bar{\nu}_L & \bar{N}^c_R
    \end{pmatrix}
    \begin{pmatrix}
    0 & m_D^T\\
    m_D & M_R
    \end{pmatrix}
    \begin{pmatrix}
    \nu^c_L \\ N_R
    \end{pmatrix}\,,
\end{equation}
where the Dirac mass $m_D= y v$ is assumed to be much smaller than $M_R$. After diagonalizing, the lighter mass eigenstate is predominantly an active neutrino interaction eigenstate, with a mass of
\begin{equation}
\label{eq:Weyl_mnu}
    m_\nu \simeq \frac{y^2 v^2}{M_R}\,,
\end{equation}
which is much smaller than the neutrino Dirac mass due to the large hierarchy assumed between $M_R$ and $m_D$, while the heavier mass eigenstate is predominantly the singlet fermion $N_R$ with mass $M_N\simeq M_R$.

\begin{figure}[t]
    \centering
    \includegraphics[width=0.45\textwidth]{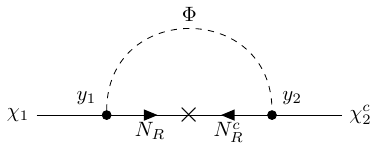}
    \caption{Lepton number violating self-energy diagram that radiatively generates a small mass splitting between the $\chi_{1,2}$ dark matter states in the type-I seesaw scenario.}
    \label{fig:mass_mixing}
\end{figure}

The lepton number breaking caused by the right-handed neutrino Majorana mass is transmitted not only to the active neutrinos via the seesaw mechanism, but also to the two dark matter particles $\chi_1$ and $\chi_2$ via quantum effects. Concretely, the diagram shown in Fig.~\ref{fig:mass_mixing} induces a contribution to the lepton number breaking part of the Lagrangian of the form
\begin{equation}
\mathcal{L}^{\cancel L}_\text{mass}\supset -A_L \left( \bar{\chi}_{1,L}^c \chi_{2,L} +\bar{\chi}_{2,L}^c\chi_{1,L} \right)\,,
\label{eq:L-quantum-seesaw}
\end{equation}
with $A_L$ given by (see Appendix~\ref{sec:app_mass_splitting} for details)
\begin{align}
    \label{eq:WeylALgeneral}
    A_L = &\frac{M_N}{(4\pi)^2} y_1 y_2\, f(M_N, M_\phi,M)\,,
\end{align}
where $f(M_N, M_\phi,M)\sim\mathcal{O}(1)$ is a loop factor. In the physically interesting limit where  $M < M_\phi \ll M_N$, this expression can be approximated by
\begin{equation}
\label{eq:WeylAL}
    A_L \simeq \frac{M_N}{(4\pi)^2} y_1 y_2 \left(1 - 2\log\left(\frac{M_N}{\mu}\right)\right)\,,
\end{equation}
with $\mu$ being the renormalization scale, which we choose to be $\mu = M_N$. The correction is thus directly proportional to $M_N$ as well as $y_1y_2$.\footnote{Note that in the  regime $M_N \sim M_\phi$, a cancellation occurs in the loop factor which can generate small corrections even without a small prefactor in Eq.~\eqref{eq:WeylALgeneral}. However, we do not consider this possibility and instead focus on the more natural regime where the small Majorana mass and small couplings drive the mass splitting to low values.}

The radiatively generated term in Eq.~\eqref{eq:L-quantum-seesaw} leads to a tiny misalignment between the interaction eigenstates $\chi_1$ and $\chi_2$ and their mass eigenstates, which translates into a non-vanishing mass splitting $\epsilon$ (as defined in Eq.~\eqref{eq:defepsilon}). The mass splitting is approximately given by
\begin{equation}
\label{eq:epsilon_seesaw}
    \epsilon \approx \frac{A_L}{M} \approx 0.006 \,y_1 y_2 \,\left(\frac{M_N}{M}\right)\,.
\end{equation}
 
As a result, $\chi_2$ is kinematically allowed to decay into $\chi_1$ along with other particles. Again, focusing on the physically interesting limit where  $M < M_\phi \ll M_N$, one finds an effective operator of the form Eq.~\eqref{eq:Lagrangian} (and ultimately Eq.~\eqref{eq:Leff}) upon integrating-out the heavy particles. Hereby, the effective scale $\Lambda$ is calculable in terms of the fundamental parameters of the model
\begin{align}
    \label{eq:Lambda_UV}
    \frac{1}{\Lambda^4} &\simeq \frac{y_1 y_2 y^2}{M_N^2 M_\phi^2} = \frac{y_1 y_2 m_\nu}{M_N M_\phi^2 v^2}\,,
\end{align}
where in the last step we have used Eq.~\eqref{eq:Weyl_mnu}. Replacing in Eq.~\eqref{eq:Lambda} one finds a lifetime of
\begin{align}
    \label{eq:seesaw_tau}
    \tau &\simeq \frac{5}{\pi} \left(\frac{16\pi^2}{y_1 y_2}\right)^7 \frac{M_\phi^4}{m_\nu^2 M_N^3} \nonumber \\
    &\approx 2.6\cdot10^{19} \s  \cdot \left(\frac{0.1}{y_1 y_2}\right)^{7} \left(\frac{M_\phi}{10\,\GeV}\right) \left(\frac{M_\phi}{M_N}\right)^{3}
    \left(\frac{0.1\eV}{m_\nu}\right)^2\,,
\end{align}
or alternatively Eq.~\eqref{eq:app_tau_S}. Following~\cite{Fuss:2024dam}, we assume (sub-)GeV masses for $\chi_1$ and $\chi_2$, as well as $\epsilon\ll 1$. Since the couplings $y_{1,2}$ are bounded from above, the lifetime falls within the cosmologically interesting regime for mediator masses $\sim 1-30 \GeV$. 

Due to the lepton number breaking, other decay channels become possible, such as $\chi_2\rightarrow \chi_1\gamma$ or $\chi_2\rightarrow \chi_1 e^+ e^-$. These channels will be discussed in Sec.~\ref{sec:L_violation_decays}.

\subsection{Inverse seesaw scenario}
\label{sec:inverse_seesaw}

In this model, we add a singlet fermion with left-handed chirality, $S_L$, to the particle content of the seesaw model. The quantum numbers of the different particles under the global symmetries $U(1)_\chi$ and $U(1)_L$ are then given by:
\begin{center}
    \begin{tabular}{c|c|c|c|c|c}
         & $\chi_2$ & $\chi_1$ & $N_R$ & $S_L$& $\phi$\\
         \hline
        $U(1)_\chi$ & +1 & $-1$ & 0 & 0 & +1\\
         \hline
        $U(1)_L$ & +1 & +1 & +1 & +1 & 0\\
    \end{tabular}
\end{center}
With this assignment, the Yukawa Lagrangian, as well as the lepton number conserving and the lepton number violating mass term Lagrangians, contain the following new terms
\begin{align}
    & \mathcal{L}_\text{Yukawa} \supset  y'_1\bar S_L \chi_1 \phi + y'_2\bar{\chi_2} S_L \phi + {\rm h.c.}\,,\nonumber \\
    & \mathcal{L}^{L}_\text{mass} \supset -M_{LR} \bar S_L N_R\,+{\rm h.c.}\,,\nonumber \\
    & \mathcal{L}^{\cancel L}_\text{mass} \supset -\frac{1}{2}\mu_S \bar S_L^c S_L \,.
\end{align}
The mass matrix is generally given by
\begin{equation}
    \begin{pmatrix}
    \bar{\nu}_L & \bar{N}^c_R & \bar{S}_L
    \end{pmatrix}
    \begin{pmatrix}
    0 & m_D & 0\\
    m_D & M_R & M_{LR}\\
    0 & M_{LR} & \mu_S
    \end{pmatrix}
    \begin{pmatrix}
    \nu^c_L \\ N_R \\ S^c_L
    \end{pmatrix}\,,
\end{equation}
with $m_D= y v$. Assuming a mass hierarchy $M_{R}\ll \mu_S\ll m_D \ll M_{LR}$, so that the lepton number breaking is small and is dominated by $\mu_S$, one finds that the lightest neutrino mass eigenstate has a mass of
\begin{equation}
\label{eq:pD_mnu}
    m_\nu \sim \frac{y^2 v^2 \mu_S}{M_{LR}^2}\,,
\end{equation}
and is dominated by the active interaction eigenstate. On the other hand, the two heavier neutrino mass eigenstates form a pseudo-Dirac pair with mass $M_N\simeq M_{LR}$ which is mostly constituted by the singlet fermions $N_R$ and $S_L$.
\begin{figure}[t]
    \centering
    \includegraphics[width=0.45\textwidth]{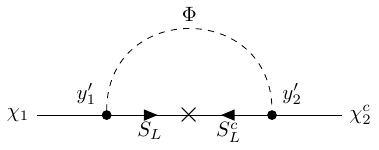}
    \includegraphics[width=0.45\textwidth]{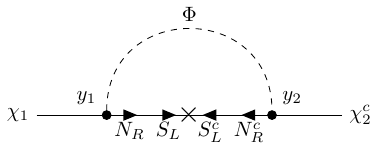}
    \caption{Same as Fig.~\ref{fig:mass_mixing} but for the inverse seesaw scenario. }
    \label{fig:mass_mixing_invS}
\end{figure}

Similarly to the seesaw case discussed in the previous subsection, the lepton number breaking is transmitted to the dark sector particles $\chi_1$ and $\chi_2$ via quantum effects as shown in Fig.~\ref{fig:mass_mixing_invS}. In the left diagram, this takes place by a direct coupling to $S_L$ that experiences the Majorana flip, while on the right side there is first a coupling to $N_R$ which is connected to $S_L$ via $M_{LR}$. The lepton number breaking part of the Lagrangian then receives a contribution of the form
\begin{equation}
\mathcal{L}^{\cancel L}_\text{mass}\supset - A_R \left( \bar{\chi}_{1,R}^c \chi_{2,R} +\bar{\chi}_{2,R}^c\chi_{1,R} \right),
\label{eq:L-quantum-inverse}
\end{equation}
with
\begin{align}
    \label{eq:invSeesaw_generalAR}
    A_R = &\frac{\mu_S}{(4\pi)^2} \left(y'_1 y'_2\, f'(M_N, M_\phi,M) +y_1 y_2\, f(M_N, M_\phi,M)\right)\,,
\end{align}
and loop factors $f,f'$. Modifications to the kinetic term of $\chi_i$ also appear, however, they are suppressed with $M/M_N$ and can be neglected (see Appendix~\ref{sec:app_mass_splitting}). 

In the limit $M<M_\phi \ll M_N$, $A_R$ can be approximated by
\begin{align}
    \label{eq:pDAR}
    A_R \simeq &\frac{\mu_S}{(4\pi)^2} \left(y'_1 y'_2 \left( -\frac{1}{2} + 2\log\left(\frac{M_N}{\mu}\right)\right)+y_1y_2\right)\,,
\end{align}
where $\mu$ is again the renormalization scale that we choose to be $\mu=M_N$. Thus, the mass splitting parameter $\epsilon$ is given by
\begin{equation}
\label{eq:eps_inverseSeesaw}
    \epsilon \simeq \frac{A_R}{M} \approx 0.006\, \delta \,\left(y_1 y_2-\frac{1}{2}y'_1 y'_2\right) \,
    \left(\frac{M_N}{M}\right)\,,
\end{equation}
where we have introduced the small parameter
\begin{equation}
\label{eq:delta}
    \delta=\frac{\mu_S}{M_N}\,.
\end{equation}

Notice that the radiatively generated mass splitting must be proportional to the order parameter of the lepton number breaking. In the seesaw scenario, this order parameter is given by $M_N\gg m_D$, while in the inverse seesaw it is $\mu_S\ll m_D$. Therefore, for comparable Yukawa couplings in both models, the mass splitting between $\chi_1$ and $\chi_2$ will typically be smaller in the inverse seesaw model than in the type I seesaw model, concretely by a factor $\sim \delta$, as apparent from comparing Eqs.~\eqref{eq:eps_inverseSeesaw} and \eqref{eq:epsilon_seesaw}.

Integrating out both the heavy scalar $\phi$ and the heavy fermions $N_R$ and $S_L$ one finds an effective coupling with
\begin{equation}
    \label{eq:Lambda_UV_inv}
    \frac{1}{\Lambda^4} \simeq \frac{y_1 y_2 y^2}{M_{LR}^2 M_\phi^2} \simeq \frac{y_1 y_2 m_\nu}{\mu_S M_\phi^2 v^2} = \delta \frac{y_1 y_2 m_\nu}{M_N M_\phi^2 v^2}\,.
\end{equation}
After combining with Eqs.~\eqref{eq:Lambda} and \eqref{eq:eps_inverseSeesaw}, this leads to a lifetime of
 \begin{align}
    \label{eq:invseesaw_tau}
    \tau &\simeq \frac{5}{\pi} \frac{\left(16\pi^2\right)^7}{y_1^2 y_2^2\left(y_1y_2-\frac{1}{2}y_1'y_2'\right)^5} \frac{M_\phi^4}{m_\nu^2 \mu_S^3}\,,
\end{align} 
or alternatively Eq.~\ref{eq:app_tau_invS}. For simplicity, we consider in what follows $y_1y_2=y_1'y_2'$  so that 
 \begin{align}
    \label{eq:invseesaw_tau_num}
    \tau &\approx 8.2\cdot10^{20} \s  \,
    \left(\frac{0.1}{y_1y_2}\right)^7
    \delta^3 \left(\frac{M_\phi}{10\,\GeV}\right) \left(\frac{M_\phi}{M_N}\right)^{3} \left(\frac{0.1\eV}{m_\nu}\right)^2\,,
    \end{align}
with $\delta$ defined in Eq.~\eqref{eq:delta}. The resulting lifetime is a factor $\sim \delta^3$ smaller than in the type-I seesaw scenario, thus allowing the possibility of larger $N$ and $\phi$ masses for the same lifetimes.

\section{Observational consequences and benchmark models}
\label{sec:benchmarks}

The models discussed in the previous section have observational signals both from cosmological surveys and from the detection of a diffuse neutrino flux, as discussed in Sec.~\ref{sec:MinimalDCDM}, and also in laboratory experiments, from the mixing of the active neutrinos and the singlet fermions, which act as heavy neutral leptons (HNLs).

The $\epsilon-\tau$ parameter space of our decaying dark matter model is shown in   Fig.~\ref{fig:Benchmarks}, along with  four benchmark points (see below). 
The gray regions at the bottom show the excluded regions due to cosmological signatures arising from the velocity kick imparted on the daughter particle $\chi_1$ in the decay (see~\cite{Fuss:2024dam} for details), where the hatched region corresponds to a Bayesian analysis based on weak lensing data from KiDS~\cite{Bucko:2023eix}. A more recent frequentist approach including the same data set showed weaker constraints~\cite{Montandon:2025xpd} and thus points to potential volume effects in the previous analysis which makes their constraints more uncertain.
The olive contours on the right show constraints on the diffuse neutrino (and anti-neutrino) flux generated via the decays $\chi_2\to\bar \chi_1\nu\nu$ (and $\bar \chi_2\to \chi_1\bar\nu\bar\nu$) from various neutrino detectors (see~\cite{Fuss:2024dam}), with the dashed line showing future sensitivity of JUNO~\cite{ColomerMolla:2023ppf,Akita:2022lit}. 
\begin{figure} [t]
    \centering
    \includegraphics[width=0.75\textwidth]{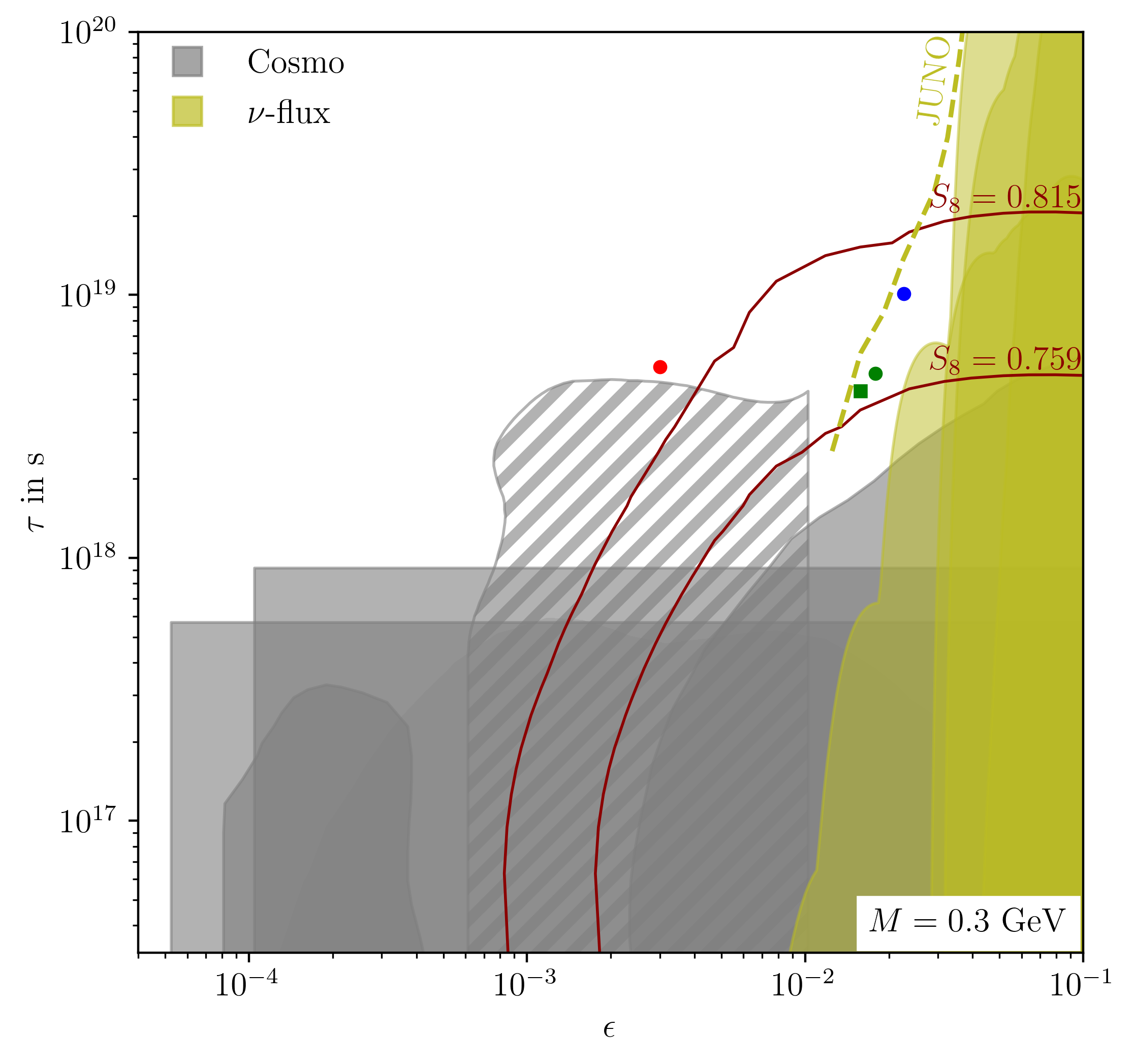}
    \caption{Lifetime $\tau$ of the DM decay $\chi_2\to\bar \chi_1\nu\nu$ versus mass splitting $\epsilon$ of the decaying DM particles $\chi_{1,2}$, assuming DM mass $M=0.3\,\GeV$. Olive regions show constraints on the diffuse neutrino (and equally strong antineutrino) flux from DM decay from various neutrino experiments (see~\cite{Fuss:2024dam} for details), and the dashed line shows a forecast for JUNO. Gray regions are excluded by cosmological constraints summarized in Sec.~\ref{sec:MinimalDCDM}. The hatched region corresponds to Bayesian weak lensing results (see text for details). The benchmark points for the inverse seesaw case (see Table~\ref{tab:Benchmarks}) are marked with circles while the seesaw case is shown with the square. In both cases, interesting parts of the parameter space can be reached, including (slightly) reduced $S_8$ values testable with large-scale structure surveys, as well as a diffuse (anti-)neutrino flux potentially detectable by JUNO.}
    \label{fig:Benchmarks}
\end{figure}

 One of the signatures of DM decay is a suppression of the power spectrum, leading to a lower value of $S_8$  measured by large-scale structure observations as compared to the one inferred from CMB data under the hypothesis of $\Lambda$CDM. The dark-red lines indicate lines where $S_8=0.759$ being the value obtained from weak lensing shear measurements by KiDS~\cite{KiDS:2020suj} in 2020, producing the strongest $S_8$ tension within $\Lambda$CDM, as well as $S_8=0.815$~\cite{Wright:2025xka} from the KiDS-Legacy analysis in 2025 where more data were considered and different analysis techniques applied. While this newer result is statistically compatible with CMB/$\Lambda$CDM based values of $S_8$ (\emph{e.g.}~$S_8=0.830\pm0.013$ for Planck~\cite{Planck:2018vyg}), the trend of a suppressed power spectrum on smaller scales is still present~\cite{Abdalla:2022yfr,Doux:2025vru} in other weak lensing experiments like DES~\cite{DES:2021wwk} or HSC~\cite{Sugiyama:2023fzm,2024AstHe.117..304S}, and in analyses based on cluster counts (\emph{e.g.}~\cite{SPT:2018njh}) and galaxy clustering (\emph{e.g.}~\cite{Ivanov:2019pdj}) \textendash~albeit smaller than from the earlier KiDS analysis. Thus, we still highlight the region between the old and new KiDS result to give an indication where a detectable amount of power spectrum suppression occurs due to DM decay.

\begin{figure}[t]
    \centering
    \includegraphics[width=0.75\textwidth]{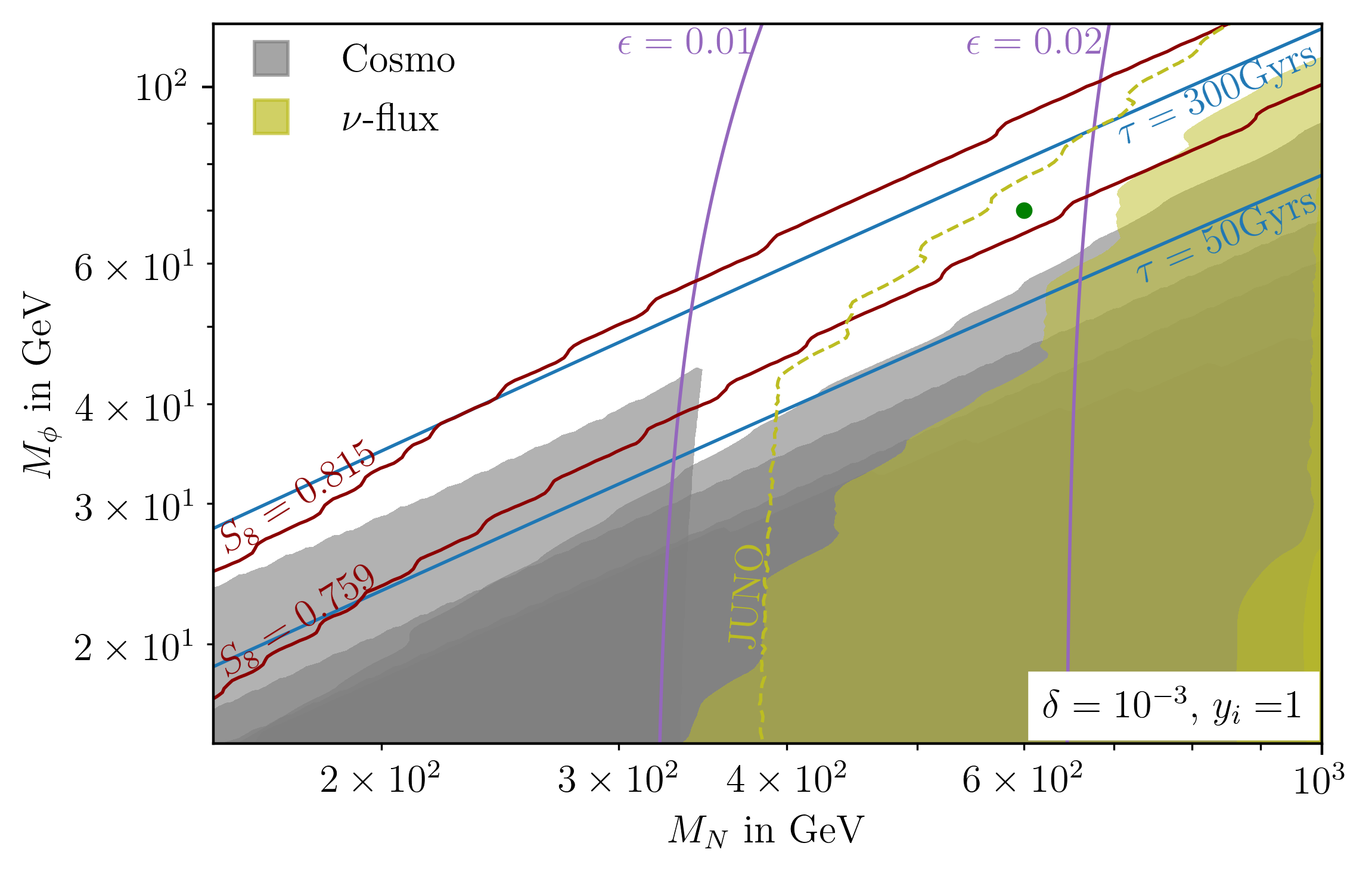}\\
    \includegraphics[width=0.75\textwidth]{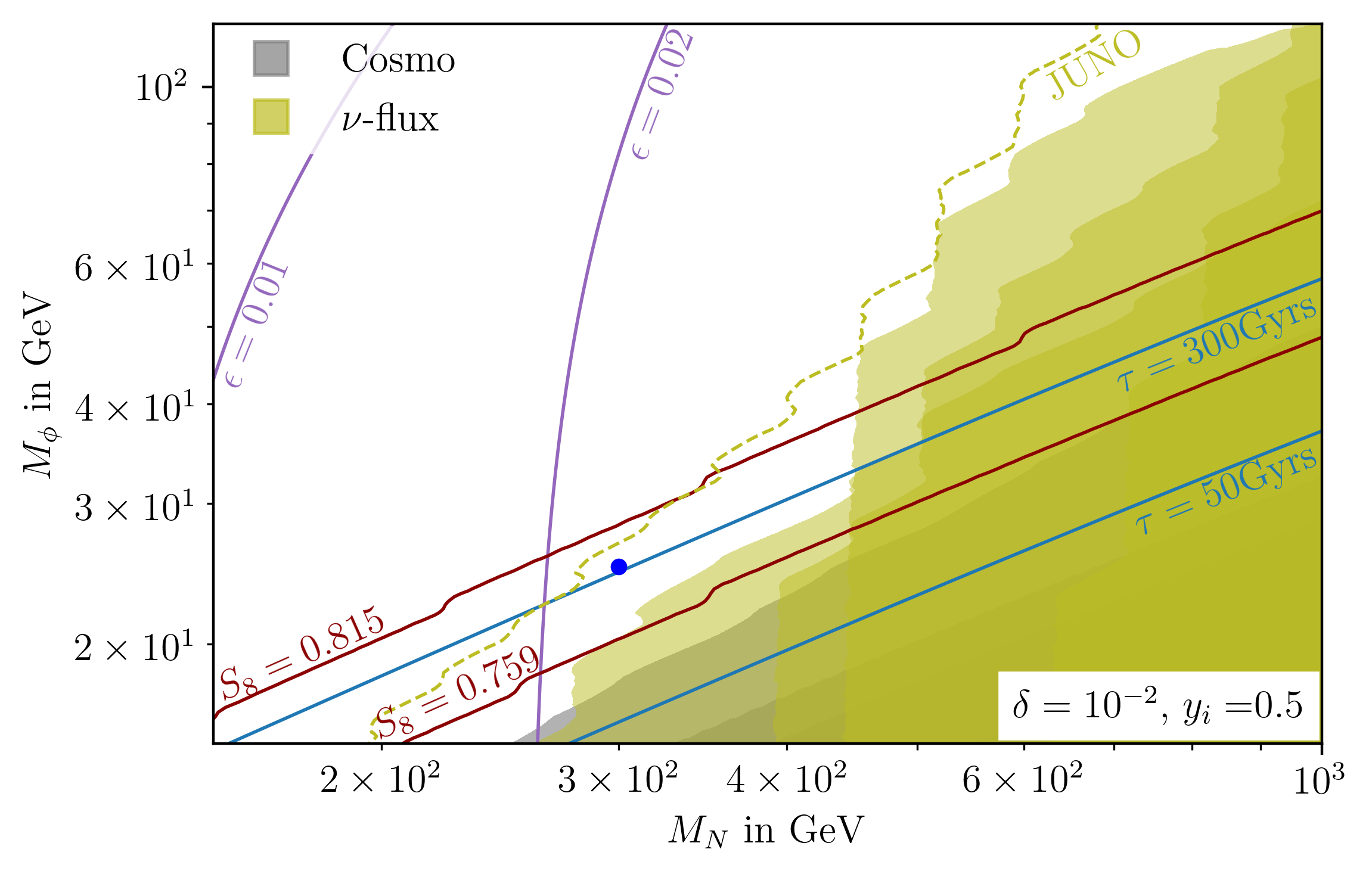}
    \caption{Same as Fig.~\ref{fig:Benchmarks}, but in the parameter space spanned by the HNL mass $M_N$ and the scalar mediator mass $M_\phi$, when $\delta = 10^{-3}$ and $y_i=1$ (top panel) and when $\delta = 10^{-2}$ and $y_i=0.5$ (bottom panel). We also show for reference the contours of the mass splitting parameter $\epsilon$ (purple line), the dark matter lifetime $\tau$ (light blue line) and $S_8$ (dark red line). The plots also include the location of the green circle (top panel) and the blue circle (bottom panel) benchmark points introduced in Table~\ref{tab:Benchmarks}.}
    \label{fig:pD_greenBenchmark}
\end{figure}

Alternatively, the various constraints can be shown in the parameter space spanned by the fundamental parameters of the model. In  Fig.~\ref{fig:pD_greenBenchmark} we show in particular the parameter space in terms of the masses of the heavy fermions and the heavy scalars, $M_N$ and $M_\phi$, when $\delta = 10^{-3}$ and $y_i=1$ (left panel) and when $\delta = 10^{-2}$ and $y_i=0.5$ (right panel).
The color coding for the projected cosmological and neutrino flux constraints is the same as in Fig.~\ref{fig:Benchmarks} and additional lines are shown in light blue for the lifetimes $\tau=50,300\gyr$ as well as the mass splitting $\epsilon=0.01,0.02$ in purple, encompassing a region of phenomenological interest. The mass splitting $\epsilon$ increases towards the right side due to the growing Majorana mass $\delta M_N$, and consequently the neutrino constraints are located on the right as well since neutrinos become more energetic. The lifetime increases towards the upper left corner, where the lower mass $M_N$ leads to a lower Yukawa coupling $y$ if the neutrino mass is fixed and thus to a less efficient $\chi_2$ decay. Additionally there is an interplay of $\tau$ and $\epsilon$ contours as the lifetime scales as $\tau\propto\epsilon^{-5}$. Overall, \lcdm is approached in the upper left corner such that $S_8$  (shown in  dark red) also increases towards the Planck value~\cite{Planck:2018vyg} in this direction. In contrast, the bottom right corner is the most constrained since it implies the largest deviation from standard cold DM due to relatively fast decays with large mass splittings. 

It follows from the figures that the model can lead to observable signals in cosmological observations if $\tau\lesssim 10^{19}$s when $\epsilon\sim 10^{-4}-10^{-1}$ and in neutrino experiments if $\tau\lesssim 10^{20}$s and $\epsilon \sim 2\times 10^{-2}-10^{-1}$ for DM masses $M=0.3\,\GeV$. These values of the lifetime are in the ballpark of the ones expected both in the type-I seesaw and the inverse seesaw, {\it cf.} Eqs.~\eqref{eq:seesaw_tau} and \eqref{eq:invseesaw_tau_num}, although in the inverse seesaw these lifetimes can be achieved also for larger values of $M_\phi$ and $M_N$, due to the existence of the additional parameter $\delta$. Similarly, the values of $\epsilon$ can be easily accommodated in the model, {\it cf.} Eqs.~\eqref{eq:epsilon_seesaw} and 
\eqref{eq:eps_inverseSeesaw}. 
 
We  consider for illustration four benchmark models with parameters given in Table~\ref{tab:Benchmarks}, the first benchmark corresponding to a type-I seesaw, and the other three to the inverse seesaw. In all the cases we fix the mass of $\chi_2$ to be $M=0.3\,$GeV and we assume $y_1=y_2$ for the Yukawa couplings. The last three columns of the table indicate the lifetime of $\chi_2$ and the mass splitting $\epsilon$, calculated using Eqs.~\eqref{eq:WeylALgeneral} and~\eqref{eq:invSeesaw_generalAR}, taking for definiteness $m_\nu = 0.1\,\eV$, as well as the value of $S_8$.

\begin{table}[t]
    \centering
    \begin{tabular}{c||c|c|c|c|c||c|c|c}
         & $M_N$ [GeV]  & $\delta$ & $M_\phi$ [GeV] & $y_i$ & $y$ & $\tau$ [$\gyr$] & $\epsilon$ & $S_8$ \\
         \hline
         seesaw \tikz\draw[tabgreen,fill=tabgreen] (0,0) rectangle (1ex,1ex); & 20 & - & 2 & 0.2 & $1.8\cdot 10^{-7}$ & 136.4 & 0.016 & 0.77 \\
         \hline
         inv. seesaw \tikz\draw[tabgreen,fill=tabgreen] (0,0) circle (.5ex); & 600 & 0.001 & 70 & 1 & $3.1\cdot 10^{-5}$ & 159.4 & 0.018 & 0.78 \\
         \hline
         inv. seesaw \tikz\draw[tabblue,fill=tabblue] (0,0) circle (.5ex); & 300 & 0.01 & 25 & 0.5 & $7.04\cdot 10^{-6}$ & 320.3 & 0.023 & 0.80 \\
         \hline
         inv. seesaw \tikz\draw[tabred,fill=tabred] (0,0) circle (.5ex); & 80 & 0.002 & 12 & 0.8 & $8.1\cdot 10^{-6}$ & 168.9 & 0.003 & 0.82
    \end{tabular}
    \caption{Parameter values for the benchmark models considered in this work. The last three columns indicate the lifetime $\tau$, the radiatively induced relative mass splitting parameter $\epsilon$, and the $S_8$ parameter characterizing the level of suppression of the matter power spectrum. For all benchmark points the DM mass is fixed to $M=0.3$\,GeV.}
    \label{tab:Benchmarks}
\end{table}

The benchmark model marked in green in Table~\ref{tab:Benchmarks} lies within the region of interest where relevant $\tau$ and $\epsilon$ parameters are combined with a lowered value $S_8=0.78$, as well as future testability by JUNO. As can be seen in Fig.~\ref{fig:pD_greenBenchmark}, this region of interest corresponds to HNL masses $M_N$ in the $\mathcal{O}(100)~\GeV$ range. For the scalar mediator, the most relevant mass region is $M_\phi\lesssim \mathcal{O}(10)~\GeV$.
For the blue benchmark, the larger Majorana admixture $\delta=10^{-2}$ leads to larger mass splittings for the same mass values and hence the figure is dominated by the neutrino flux constraints with the cosmological limits being barely visible at the bottom. The benchmark still features a decreased $S_8=0.80$ but to a lesser extent than before while still lying in the JUNO forecast region.

In summary, we find that there are regions of the parameter space of the model which can lead to observable signatures both in cosmological observations and in neutrino detectors, and which could furthermore explain the $S_8$ tension, should this tension be confirmed by experiments. 

\begin{figure}[t]
    \centering
    \includegraphics[width=0.75\textwidth]{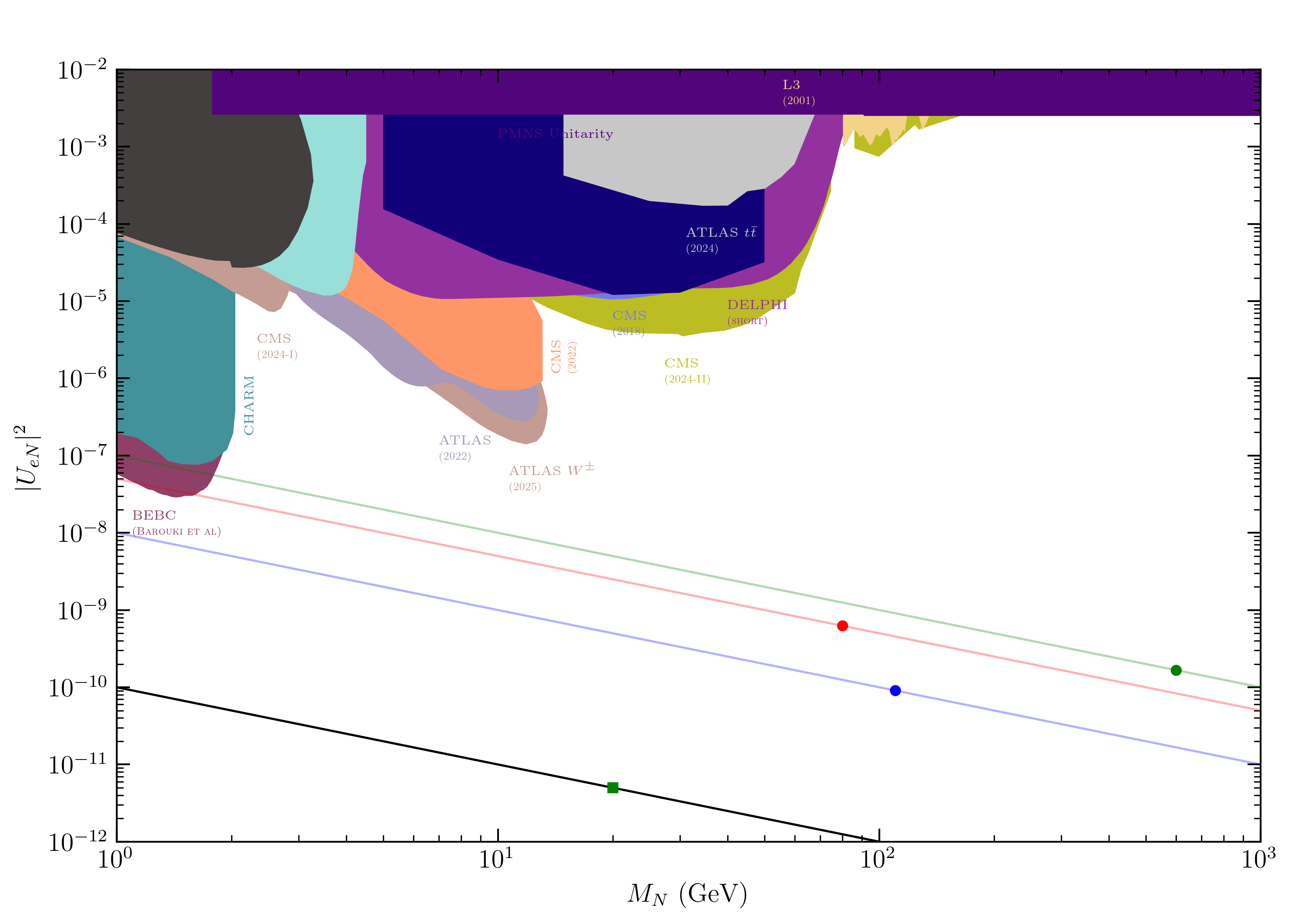}
    \caption{Laboratory constraints on the mixing of a heavy neutral lepton with the electron neutrino as a function of the mass, adopted from~\cite{Fernandez-Martinez:2023phj}. The black line indicates the prediction for the type-I seesaw models, while the red, blue and green lines for the inverse seesaw models with  $\delta\equiv\mu_S/M_N=$0.03, 0.01 and 0.001 respectively. The symbols show the benchmark points from Table~\ref{tab:Benchmarks}.}
    \label{fig:HNL_seesaw}
\end{figure}

This model could also produce signatures in laboratory experiments, due to the mixing of the heavy neutral lepton with the active neutrino, approximately given by $|U|^2\sim \frac{m_\nu}{M_M}$. The constraints from laboratory experiments, assuming that the heavy neutral lepton couples dominantly with the electron neutrinos, are shown in Fig.~\ref{fig:HNL_seesaw} as a function of the HNL mass. The shaded regions show constraints adopted from~\cite{Fernandez-Martinez:2023phj}, including laboratory probes like beam dump experiments, high-energy collider searches and electroweak precision tests (see \emph{e.g.}~\cite{Bechtol:2022koa} for a  detailed overview). 
Note that for $N$ masses below the $\GeV$ scale, additional cosmological limits based on Big Bang Nucleosynthesis (BBN) and CMB observations exclude most of the parameter space down to extremely low mixing angles and only relax at around $\text{keV}$ scales. In between, the HNL decays into SM particles~\cite{Bondarenko:2018} and the accompanying energy injection and changed expansion history will lead to signatures in \emph{e.g.}~$N_\text{eff}$, the polarization spectrum or modified element abundances~\cite{Vincent:2014rja}. Thus, we restrict our analysis to masses $M_N\gtrsim 1\,\GeV$ and show only the constraints in this range which is in agreement with the requirement of $N$ being heavier than the DM particles $\chi_{1,2}$. 
The black diagonal line shown in Fig.~\ref{fig:HNL_seesaw} marks the standard seesaw relation from Eq.~\eqref{eq:Weyl_mnu}, while the inverse seesaw mechanism opens up the parameter space above due to the well-known additional suppression of $m_\nu$ from the small Majorana mass $\mu_S$, see Eq.~\eqref{eq:pD_mnu}. In Fig.~\ref{fig:HNL_seesaw} we also indicate the benchmark models presented in Table~\ref{tab:Benchmarks}, which are clearly compatible with current laboratory constraints.
Note that apart from the HNL there are additional dark sector particles present, including the scalar mediator $\phi$. While they couple only very indirectly to the SM which makes them hard to probe, they might still affect cosmological and laboratory bounds (see~\cite{Dev:2025pru}) by opening up new dark decay channels $N\rightarrow \phi \chi_i$, and a future more detailed study on these effects would be of interest.

Finally, let us briefly comment on the generation of the observed dark matter abundance within our model. 
As is well-known, for HNL masses in the GeV-TeV range and mixing compatible with the seesaw (or inverse seesaw) relation, the thermal production of the HNL $N$ is efficient and thermalization with the SM is achieved if the reheating temperature is above the HNL mass, $T_{rh}\gg M_N$~\cite{Boyarsky:2020dzc,Ghisoiu:2014ena,Ghiglieri:2016xye} (see~\cite{Alonso-Alvarez:2022uxp,Dodelson:1993je,Dolgov:2000ew} for production mechanisms at smaller mixing parameters which however do not apply in our case). Since the coupling  between the dark sector and $N$ via the scalar mediator is not suppressed by any small parameter, \emph{i.e.}~$y_i\sim\mathcal{O}(1)$, the dark sector particles $\chi_{1,2}$ and $\phi$ consequently equilibrate with $N$, and thus also with the SM thermal bath. This potentially leads to an overproduction of $\chi_{1,2}$, unless their density is sufficiently depleted via thermal freeze-out. Since the dark sector states $N$ and $\phi$ are heavier than the $\chi_{1,2}$, no freeze-out channels within the dark sector are available. Annihilation channels into light SM states are however heavily suppressed, being related to the small HNL mixing associated with the smallness of neutrino masses. Within the model studied in this work the same suppression is also responsible for guaranteeing the cosmologically long lifetime of the decay $\chi_2\to \bar \chi_1\nu\nu$. Thus, no freeze-out channels with sufficiently large cross section are available.

We therefore consider the possibility that the reheating temperature is below the HNL mass, $T_{rh}\lesssim M_N$.
Due to the preferred mass range $M_N\sim {\cal O}(10^2-10^3)$GeV this is a viable scenario, compatible with BBN and inflationary dynamics in the Early Universe. In this case the HNL and thus the dark sector does not thermalize with the SM. Instead, as pointed out already in~\cite{Fuss:2024dam}, the DM abundance can be explained by freeze-in, via the same effective interaction Eq.~\eqref{eq:Leff} that is also responsible for the DM decay. Specifically, it gives rise to the freeze-in process $\nu\nu\rightarrow  \chi_1  \chi_2$, with a strength that matches the observed DM density if the reheating temperature is in the GeV\textendash100~GeV range.
Following~\cite{Fuss:2024dam}, we obtain a reheating temperature $T_{rh}\sim 70\,\GeV$, $T_{rh}\sim 150\,\GeV$ and $T_{rh}\sim 5\,\GeV$ for the green, blue and red benchmark model, respectively. These values are all well below the respective $M_N$ values (see Table~\ref{tab:Benchmarks}), being in accord with the assumed absence of thermal HNL production. Furthermore, we note that additional production processes for freeze-in are possible via the box diagrams including lepton violation discussed in Sec.~\ref{sec:app_L_violating_decays}. We checked that these are either subdominant or at maximum of the same order in the required temperature range. These findings present a non-trivial consistency check of the EFT description of DM freeze-in production within the UV completion studied in this work, and demonstrate that it is a viable scenario in this setup. 

\section{Lepton number violating decay channels}
\label{sec:L_violation_decays}

\begin{figure}[t]
    \centering
    \includegraphics[width=0.349\textwidth]{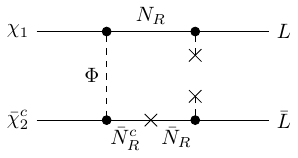}
    \hspace{1cm}
    \includegraphics[width=0.349\textwidth]{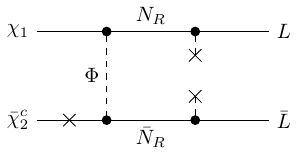}
    \caption{\label{fig:box_diagram_L_violation}Examples of diagrams in the seesaw case that induce subleading decay modes $\chi_2 \to\bar \chi_1 e^+ e^-$ via soft lepton number breaking associated to the HNL Majorana mass responsible for neutrino mass generation. The lepton number violation can either enter via the propagator (left) or one of the external legs (right) due to the mass mixing between $\chi_1$ and $\chi_2$, as indicated by the cross, respectively.}
\end{figure}

In this section we will investigate the implications of lepton number violating decays with charged particles or photons in the final state, and which could be detected in cosmic-ray or gamma-ray detectors. More concretely, in the presence of lepton number violation, the one loop diagrams shown  in Fig.~\ref{fig:box_diagram_L_violation} generate the dimension-six operator $\mathcal{L}^{(6)}_{\cancel L}\sim \left(\chi_1 \bar L \right) \left(\bar \chi_2^c L \right)$,\footnote{Another possible  box diagram producing an $e^\pm$-pair is obtained by replacing the Higgs propagator in the left diagram in Fig.~\ref{fig:box_diagram_L_violation} by the insertion of two Higgs-VEVs, and adding a $W$ exchange (see Fig.~\ref{fig:W_exchange} in Appendix~\ref{sec:app_L_violating_decays}). This diagram gives a contribution which is numerically comparable to the one from the diagrams in Fig.~\ref{fig:box_diagram_L_violation}.} which in turn leads to the decay $\chi_2\rightarrow \bar \chi_1 e^+ e^-$.
    
For the seesaw case, the lepton number violation via the $N_R$ propagator is unsuppressed (as far as the propagator line is concerned) since it is a pure Majorana state with $N=N^c$. In Appendix~\ref{sec:app_L_violating_decays} we estimate the effective scale of this process and compare to the neutrino decay, yielding a branching ratio
\begin{align}
\label{eq:seesaw_BR_leptonviol}
    \frac{\Gamma_{\chi_2\rightarrow \bar\chi_1 e^+ e^-}}{\Gamma_{\chi_2\rightarrow \bar \chi_1 \nu \nu}} \sim 
    \frac{M_N^4 M_\phi^4}{\left(16\pi^2 v^2\right)^2 M_h^4 } \approx \,1.1\cdot 10^{-16}\left(\frac{M_N}{20\,\GeV}\right)^4 \left(\frac{M_\phi}{10\,\GeV}\right)^4\,,
\end{align}
where $M_h$ is the Higgs mass and we apply typical mass hierarchies. For the ranges of parameters leading to cosmological and neutrino signals, the branching ratio of this decay mode is at least seven orders of magnitude smaller than the sensitivity of current instruments~\cite{Ibarra:2013zia}. In the inverse seesaw case, we find instead 
\begin{align}
    \label{eq:Lvioldecays_pD}
    \frac{\Gamma_{\chi_2\rightarrow \bar\chi_1 e^+ e^-}}{\Gamma_{\chi_2\rightarrow \bar \chi_1 \nu \nu}} \sim &\frac{\delta^2 M_N^2 M_\phi^4}{\left(16\pi^2 v^2\right)^2 M_h^2} \nonumber \\
    \approx&\, 6.1\cdot 10^{-12} \left( \frac{\delta}{0.001}\right)^2 \left( \frac{M_N}{600\,\GeV}\right)^2 \left( \frac{M_\phi}{70\,\GeV}\right)^4\,,
\end{align}
which is still far below the sensitivity of current instruments. 

The electron-positron-pairs in Fig.~\ref{fig:box_diagram_L_violation} can also be connected to a loop, to which a photon is attached, leading to a monochromatic gamma-ray line with energy $E_\gamma = \epsilon M$ via the decay $\chi_2\to\bar \chi_1\gamma$ at two-loop order. While indirect detection constraints are stronger in this case, the needed additional loop also leads to a further suppression~\cite{Garny:2010eg}, making this channel similarly (un-)constraining for the decaying DM model under consideration. Nevertheless, this decay channel may comprise an interesting possibility if such a gamma-ray line in the MeV range is detected in the future, for example by the planned space-based COSI telescope~\cite{Tomsick:2019wvo,Beechert:2022phz}.

\section{Conclusions}
\label{sec:conclusion}

We have constructed a model of dark matter which naturally links the existence of non-zero neutrino masses to dark matter decay, and which could produce signatures both in cosmology and in neutrino experiments. The model involves two singlet fermions $\chi_1$ an $\chi_2$ degenerate in mass and with identical lepton number, but opposite charges under an additional global symmetry. If the lepton number is exactly conserved, both $\chi_1$ and $\chi_2$ are absolutely stable. However, if the lepton number is broken by two units, non-zero neutrino masses can be generated, a mass splitting between $\chi_1$ and $\chi_2$ can be induced by quantum effects, and the heavier dark matter component can decay into the lighter and two neutrinos. 

We have considered two possible scenarios for lepton number violation motivated by archetype models of neutrino mass generation: the type-I seesaw mechanism and the inverse seesaw mechanism. For each of these scenarios we have calculated the mass splitting between $\chi_1$ and $\chi_2$ at the one-loop level, as well as the decay rate of the process $\chi_2\rightarrow  \bar \chi_1 \nu\nu$. Notably, we have found regions in the parameter space which can lead to signatures in cosmological surveys and in neutrino experiments. Concretely, the daughter particle acquires in the decay a slight ``kick" that suppresses the matter power spectrum, thereby leaving an imprint in the large-scale structure of the Universe, and in particular sizably modify the $S_8$ parameter. Furthermore, the decay generates a diffuse neutrino flux that could be detected at Super-Kamiokande, KamLAND or the upcoming JUNO experiment. We have also found that the parameter space that can lead to this signatures is larger for the inverse seesaw scenario than for the type-I seesaw scenario, due to the smaller order parameter of the lepton number symmetry, which allows for larger masses for the heavy particles mediating the decay.

Finally, we have briefly discussed the possibility of detecting in experiments lepton number violating decay modes, such as $\chi_2\rightarrow \bar \chi_1 e^+ e^-$ or $\chi_2\rightarrow \bar \chi_1 \gamma$, as well as possible mechanisms of dark matter production. 

\subsection*{Acknowledgements}

We acknowledge support by the DFG Collaborative Research Institution Neutrinos and Dark Matter in Astro- and Particle Physics (SFB 1258) and the Excellence Cluster ORIGINS - EXC-2094 - 390783311.

\clearpage
\appendix

\section{Natural mass degeneracy from a \texorpdfstring{\boldmath$SU(2)$}{SU(2)} symmetry}
\label{sec:app_symmetries}

\begin{figure}[t]
    \centering
    \includegraphics[width=0.562\textwidth]{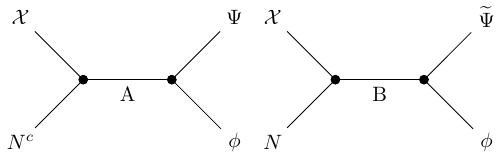}
    \caption{Diagrams responsible for the Yukawa couplings between the DM particles $\chi_{1,2}$, the scalar mediator $\phi$ and the HNL $N$, within the extended dark sector setup featuring dark isospin symmetry.}
    \label{fig:UV2_Yukawas}
\end{figure}

In this appendix we outline an extension of the symmetry group describing the dark sector, for which the two DM states $\chi_1$ and $\chi_2$ are part of a single multiplet. This embedding is of interest, since it provides a natural explanation for their mass degeneracy (apart from the radiatively induced splitting discussed in the main text), see Eq.~\eqref{eq:Lmass}.
Specifically, we assume an $SU(2)$ symmetry group in the dark sector, such that the $\chi_{1,2}$ states are part of a doublet. 
This symmetry can be considered as a `dark isospin'.
We consider the particular form
\begin{equation}
    \mathcal{X}=\begin{pmatrix}
    \chi_2 \\
    \chi_1^c
\end{pmatrix}\,,
\end{equation}
such that both components transform uniformly under the $U(1)_\chi$ symmetry, see Sec.~\ref{sec:Neutrino_masses}.
Dark isospin symmetry allows only a common mass term $M\overline{\mathcal{X}} \mathcal{X}$, providing a symmetry argument for the desired mass degeneracy. 
Since the two components of the doublet carry opposite lepton number it is possible to associate lepton number with dark isospin. Indeed, this suggests a construction that is analogous to the emergence of electric charge from weak isospin and hypercharge in the SM. To realize this possibility, we introduce an additional $U(1)_Z$ symmetry in the dark sector, that plays the role of hypercharge in the SM, as well as a dark Higgs field $\Psi$ that is a doublet under dark isospin and carries non-zero charge under $U(1)_Z$, such that spontaneous symmetry breaking in the dark sector gives rise to the emergence of lepton number,
\begin{equation}
    SU(2) \times U(1)_Z \times U(1)_\chi \longrightarrow U(1)_L \times U(1)_\chi\,,
\end{equation}
with charge assignments for $\mathcal{X}$ and $\Psi$ given in Table~\ref{tab:UV2}. This gives rise to a relation between lepton number, dark hypercharge $Z$ and dark isospin $I_3$ analogously as in the SM,
\begin{equation}
     L = Z/2 + I_3\,.
\end{equation}
The $\mathcal{X}$ doublet has $Z=0$ such that the two components feature $L=I_3=\pm 1$ and the model discussed in the main text is recovered after symmetry breaking.
All SM particles are singlets under dark isospin, $I_3=0$, such that their dark hypercharge is identical to their lepton number (up to a factor two), \emph{i.e.}~$U(1)_Z$ and $U(1)_L$ are indistinguishable when only SM fields are involved. The desired breaking pattern requires to assign $Z=2$ to the dark Higgs, such that its lower component (which receives a VEV) has $L=0$ (leaving $U(1)_L$ unbroken) and the upper component $L=2$.
\begin{table}[t]
    \centering
    \begin{tabular}{c|c|c|c|c|c|c}
         & $\mathcal{X}=\begin{pmatrix} \chi_2 \\ \chi_1^c \end{pmatrix}$ & $\Psi = \begin{pmatrix} \psi^{++} \\ \psi^0 \end{pmatrix}$ & N & $\phi$ & A & B \\ \hline
       $SU(2)$ & \underline 2 & \underline 2 & \underline 1 & \underline 1 & \underline 1 & \underline 1 \\
       $U(1)_Z$ & 0 & 2 & 2 & 0 & 2 & -2 \\
       $U(1)_\chi$ & 1 & 0 & 0 & 1 & 1 & 1\\ \hline
       $U(1)_L$ & 
       $\begin{cases} +1\\-1 \end{cases}$ &
       $\begin{cases} 2\\0 \end{cases}$ & 1 & 0 & 1 & -1
    \end{tabular}
    \caption{Fields and their symmetries in an extension of the dark sector symmetry group, featuring a `dark isospin' group $SU(2)$ as well as a `dark hypercharge' $U(1)_Z$ in addition to the $U(1)_\chi$ considered already in the main text. Spontaneous symmetry breaking by the dark Higgs $\Psi$ gives rise to lepton number in this setup, $SU(2) \times U(1)_Z  \longrightarrow U(1)_L$. All SM fields are singlets under dark isospin, such that lepton number is equivalent to $U(1)_Z$ for SM particles.}
    \label{tab:UV2}
\end{table}
We refrain from exploring this enhanced symmetry in more detail in this work, but provide a sketch of some of the elements that are required to reproduce the model considered in the main text.
In particular, the scalar mediator $\phi$ carries only $\chi$-charge but no lepton number. This can be realized in the enhanced symmetry by taking it to be a singlet under dark isospin and hypercharge, see Table~\ref{tab:UV2}. However, in this case the desired Yukawa couplings Eq.~\eqref{eq:Yukawa} are incompatible with the full symmetry of the model. Nevertheless, they can be generated effectively after symmetry breaking in the dark sector via the dimension-five operator
\begin{align}
    \frac{1}{\Lambda} &\overline{\mathcal{X}}\Psi N^c \phi + h.c. \longrightarrow \frac{v_\psi}{\Lambda} \bar \chi_1 N \phi^\dagger + h.c.\,,\nonumber\\
    \frac{1}{\Lambda} &\overline{\mathcal{X}} \tilde\Psi N \phi + h.c. \longrightarrow \frac{v_\psi}{\Lambda} \bar \chi_2 N \phi + h.c\,,
\end{align}
where $v_\phi$ denotes the VEV of the lower component of $\Psi$, and we made use of  the conjugate field $\tilde\Psi \equiv i \sigma^2 \Psi^*$ which conserves the isospin $I_3$ while flipping the $U(1)$ charges. The scale $\Lambda$ can be associated to the mass of some heavier particles in the dark sector. We show an example in Fig.~\ref{fig:UV2_Yukawas}, with heavy dark sector particles $A, B$ carrying charges as shown in Table~\ref{tab:UV2}.

We finally note that, in this setup, a small mass splitting between the $\chi_{1,2}$ states could also be generated as a consequence of small isospin breaking, broadly speaking analogous to the neutron-proton mass difference in the SM. This effect can be described by possible dimension-five operators
\begin{align}
    \frac{1}{\Lambda'} &\overline{\mathcal{X}}\Psi \mathcal{X}\Psi^\dagger + h.c. \longrightarrow \frac{v_\psi^2}{\Lambda'} \bar \chi_1 \chi_1 + h.c.\,,\nonumber\\
    \frac{1}{\Lambda'} &\overline{\mathcal{X}} \tilde\Psi\mathcal{X}\tilde\Psi^\dagger + h.c. \longrightarrow \frac{v_\psi^2}{\Lambda'} \bar \chi_2 \chi_2 + h.c\,,
\end{align}
leading to a small mass splitting $\epsilon\sim \mathcal{O}(v_\psi^2/(M\Lambda'))$, which depends on the effective scale $\Lambda'$. We note that the particles $A, B$ considered above would not induce such a dimension-five operator at tree-level. Therefore, it is well possible that $\Lambda'\gg \Lambda$, such that the isospin-induced mass splitting is much smaller than the radiatively induced mass splitting associated to the connection to neutrino mass generation, as discussed in Sec.~\ref{sec:Neutrino_masses} in the main text.

While this set-up is complex, it shows that it is  possible to embed an underlying symmetry in the model for explaining the common mass term for $\chi_{1,2}$ by employing known mechanisms from the SM. This extension provides further possibilities for phenomenological exploration that are left to future work.

\section{Calculation of the mass splitting}
\label{sec:app_mass_splitting}

In this Appendix we provide details on the computation of the radiatively induced mass splitting between the $\chi_{1,2}$ states discussed in Sec.~\ref{sec:Neutrino_masses}.
The one-loop self energy shown in Fig.~\ref{fig:mass_mixing} gives rise to mass correction terms stemming from $\chi_{1,\alpha}(-i\Sigma_{\alpha\beta})\chi_{2,\beta} + h.c.$. In the seesaw case, using a Majorana propagator for the internal $N_R$ line, accounting for the proper insertions of the charge conjugation operator $C$ and applying right-handed projectors $P_R$ at the vertices yields
\begin{align}
    -i \Sigma_{\alpha \beta} &= i y_1 y_2 M_N \left(C^{-1}P_L\right)_{\alpha\beta} \tilde\mu^{4-d}\int \frac{d^dk}{(2\pi)^d}\frac{1}{k^2-M_N^2+i\cdot 0} \frac{1}{k^2-M_\phi^2+i\cdot0}\\
    &= -\frac{y_1 y_2}{(4\pi)^2} M_N \left(C^{-1}P_L\right)_{\alpha\beta} \int_0^1 dx\left(\frac{2}{\varepsilon}-\log\left(\frac{\Delta(x)}{\mu^2}\right)\right)\,,
\end{align}
where $d=4-2\varepsilon$ stems from dimensional regularization, $\mu$ is the renormalization scale in the $\overline{MS}$-scheme and $\Delta(x)= -x(1-x)p^2 + (1-x)M_N^2+xM_\phi^2$. As expected, new Majorana mass terms arise 
\begin{equation}
    \chi_{1,\alpha}(-i\Sigma_{\alpha\beta})\chi_{2,\beta} = -\frac{y_1y_2}{(4\pi)^2}M_N f(M_N,M_\phi,M) \bar \chi_1^c P_L \chi_2\,\equiv -A_L \bar \chi_1^c P_L \chi_2\,,
\end{equation}
which leads to the result in Eq.~\eqref{eq:WeylAL}.

For the inverse seesaw case, a small lepton number violating mass term occurs only for the left-handed $S_L$ field
\begin{equation}
    \frac{1}{2}\mu_S \bar{S_L^c} S_L = \mu_S (C P_L)_{\alpha \beta} (S_L)_\alpha (S_L)_\beta\,,
\end{equation}
in contrast to the pure Majorana case. We take advantage of the smallness of $\mu_S$ and treat the $N$ propagator as two Dirac propagators with the small Majorana mass taken as an insertion to account for the leading effect of lepton number violation.
For both outer $\chi_{1,2}$ particles incoming, the self-energy is given by
\begin{align}
     -i\Sigma_{\alpha\beta}= &\mu^{4-d}\int \frac{d^dl}{(2\pi)^d}(-i)(y_1 P_L + y_1' P_R)_{\gamma\alpha}  \frac{i (\cancel{l} + M_N)_{\gamma'\gamma}}{l^2-M_N^2} \left(\mu_S C P_L\right)_{\gamma'\delta'}\\
     &\frac{i (-\cancel{l} + M_N)_{\delta'\delta}}{l^2-M_N^2} \frac{i}{(p-l)^2-M_\phi^2} (-i) (y_2 P_L + y_2' P_R)_{\delta\beta}\,,
\end{align}
where all couplings $y_i$ are assumed to be real and in the main text we later choose $y_i=y_i'$ for simplicity.
Due to the two inserted fermion propagators, the contributions with a momentum $\cancel{l}$ in the enumerator do not vanish as in the seesaw case and instead four contributions $f_{XY}$ arise where $X,Y$ is either the momentum or mass term from the first and respective second propagator
\begin{align}
    -i\Sigma_{\alpha\beta} = -2\mu_S \int_0^1 dx (1-x) \tilde\mu^{4-d}\int \frac{d^dl}{(2\pi)^d} \frac{\left( f_{ll} + f_{Ml} + f_{lM} + f_{MM}\right)}{\left((l-x p)^2 - \Delta(x,p^2)  + i\cdot 0\right)^3}\,.
\end{align}
Hereby, $\Delta(x,p^2) = -x(1-x)p^2 + (1-x)M_N^2 + x M_\phi^2$ and
\begin{align}
    \left(f_{ll} + f_{lM} + f_{Ml} + f_{MM}\right)_{\alpha\beta} = &- y_1' y_2' (\cancel{l}^T \gamma^2\gamma^0 \cancel{l}P_R)_{\alpha\beta}
    + y_1' y_2 M_N (\cancel{l}^T \gamma^2\gamma^0 P_L)_{\alpha\beta} \\ 
    &- y_1 y_2' M_N (\gamma^2\gamma^0 \cancel{l} P_R)_{\alpha\beta}
    + y_1 y_2 M_N^2 (\gamma^2\gamma^0 P_L)_{\alpha\beta}\,,
\end{align}
which after standard computations leads to
\begin{align}
\label{eq:SigmaInvSeesaw}
    -i\Sigma_{\alpha\beta} =& -i\mu_S(4\pi)^{-2} y_1'y_2'(\gamma^2\gamma^0 P_R)_{\alpha\beta} \nonumber\\
    & \int_0^1 \dx z_1 \left(2(1-z_1) \left(\frac{2}{\varepsilon} - \log\left(\frac{\Delta}{\mu^2}\right)  \right) + z_1^2(1-z_1)  \frac{p^2}{\Delta} \right)\nonumber\\
	& -i\mu_S(4\pi)^{-2} M_N (\cancel p ^T \gamma^2\gamma^0 \left(y_1'y_2 P_L + y_1y_2' P_R\right))_{\alpha\beta} \int_0^1 \dx z_1 z_1(1-z_1) \frac{1}{\Delta}\nonumber\\
    & -i\mu_S(4\pi)^{-2} y_1 y_2 M_N^2 (\gamma^2\gamma^0 P_R)_{\alpha\beta} \int_0^1 \dx z_1 (1-z_1) \frac{1}{\Delta}\,.
\end{align}
This is also shown in Fig.~\ref{fig:mass_mixing_invS} where the left diagram shows contributions arising due to the direct coupling to $S_L$ with $y_i'$ while the right diagram couples to $N_R$ first with $y_i$.
The first, second and last lines in Eq.~\eqref{eq:SigmaInvSeesaw} are responsible for the mass correction ($\propto \tilde A_R$) while the third line results in a wavefunction correction to the kinetic term ($\propto C_{L,R}$) instead
\begin{align}
    \mathcal{L}_{\cancel L} \supset & -\widetilde A_R \bar \chi^c_{1,R} \chi_{2,R} + C_R \bar{\chi}_{1,R}^c \cancel p \chi_{2,L} + C_L\bar{\chi}_{1,L}^c \cancel p \chi_{2,R} + h.c.\,.
\end{align}
Assuming the typical hierarchy $M\ll M_N$ considered in this work, the  $C_{L,R}$ contributions are suppressed with one power of $M/M_N$ while the contribution to $\tilde A_R$ proportional to $p^2$ in the first line of Eq.~\eqref{eq:SigmaInvSeesaw} is suppressed with $\left( M/M_N\right)^2$ when evaluating the self-energy at $p^2=M^2$. Thus, the  dominant corrections are mass terms, given by
\begin{equation}
    \mathcal{L}_{\cancel L} = -A_R \bar{\chi}_{1,R}^c\chi_{2,R} +  h.c.\,,
\end{equation}
with $A_R$ denoting the leading contribution to $\tilde A_R$ for small $M/M_N$,
\begin{align}
    A_R = &\frac{\mu_S}{(4\pi)^2} y_1' y_2' \int_0^1 \dx z_1 2(1-z_1) \left( -\frac{2}{\varepsilon} + \log\left(\frac{\Delta}{\mu^2}\right)\right) \nonumber\\
    - &\frac{\mu_S}{(4\pi)^2} y_1 y_2 M_N^2 \int_0^1 \dx z_1 (1-z_1) \frac{1}{\Delta}\,,
\end{align}
which leads to the final result in Eq.~\eqref{eq:pDAR}.

\section{Lepton number violating decay estimates}
\label{sec:app_L_violating_decays}

In this appendix, we provide some details for the loop-induced $\Delta L=2$ decay modes $\chi_2^c\to \chi_1 e^+e^-$.
We consider the box diagram shown in the left panel of Fig.~\ref{fig:box_diagram_L_violation}.
In the seesaw case, the loop integral is of the form
\begin{align}
    & y_1 y_2 y^2 M_N \mu^{4-d}\int \frac{d^dl}{(2\pi)^d} \frac{1}{M_h^2-l^2}\frac{1}{M_\phi^2-l^2} \frac{\left(\cancel l + M_N\right)_{\alpha\beta}\left(\cancel l + M_N\right)_{\gamma\delta}}{(l^2-M_N^2)^2}\,,
\end{align}
where both of the two lighter masses can be neglected without leading to an IR divergence and only the heaviest mass will contribute. 
The box diagram gives rise to a contribution to an effective operator 
\begin{equation}
    \mathcal{L}^{(6)}_{\cancel L}= \frac{1}{\Lambda^2_{\cancel{L}}} \mathcal{O}^{(6)}_{\cancel{L}} , \qquad \mbox{where} \quad \mathcal{O}^{(6)}_{\cancel{L}} \equiv\left(\chi_1 \bar L \right) \left(\bar \chi_2^c L \right)\,.
\end{equation}
The effective suppression scale $\Lambda^2_{\cancel{L}}$  is then given by
\begin{equation}
\label{eq:Lviol_Weyl}
     \mathcal{L}^{(6)}_{\cancel{L}} = \frac{1}{16\pi^2} \frac{y_1 y_2 y^2}{\text{max}(M_h,M_\phi,M_N)^2} f_\text{box}(M_\phi, M_N, M_h) \mathcal{O}^{(6)}_{\cancel{L}} \,,
\end{equation}
where we expect the exact loop factor $f_\text{box}(M_\phi, M_N, M_h)\sim \mathcal{O}(1)$. For the mass hierarchy considered in this work, $M_h$ is the largest mass involved and the branching ratio can be computed by comparing the effective scales
\begin{align}
   \frac{\Gamma_{\chi_2\rightarrow \bar \chi_1 e^+ e^-}}{\Gamma_{\chi_2\rightarrow \bar \chi_1 \nu \nu}} \approx \left(\frac{1/\Lambda^2_{\cancel{L}}}{v^2/\Lambda^4}\right)^2\,.
\end{align}
Combined with the relation
\begin{equation}
    \label{eq:app_tau_S}
    \Gamma_{\chi_2\rightarrow \bar \chi_1 \nu \nu}^{-1}\equiv \tau =\frac{1280 \pi^3}{y_1^2 y_2^2} \frac{M_\phi^4 M_N^2}{m_\nu^2\left( \epsilon M\right)^5}\,,
\end{equation} 
the estimate in the main text~\eqref{eq:seesaw_BR_leptonviol} is reproduced.

\begin{figure}[t]
    \centering
    \includegraphics[width=0.466\textwidth]{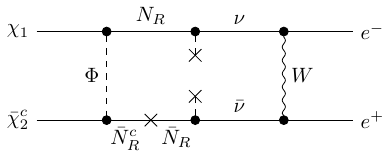}
    \caption{Additional $\Delta L=2$ process leading to an $e^\pm$-pair in the final state.}
    \label{fig:W_exchange}
\end{figure}

In the inverse seesaw case, the $L$ violating $N$ propagator inside the box diagram from the left panel of Fig.~\ref{fig:box_diagram_L_violation} is described by two Dirac propagators with a Majorana mass insertion and the loop integral we want to estimate reads
\begin{align}
    & y_1 y_2 y^2 \delta M_N \mu^{4-d}\int \frac{d^dl}{(2\pi)^d}  \frac{1}{M_h^2-l^2}\frac{1}{M_\phi^2-l^2} \frac{\left(\cancel l + M_N\right)_{\alpha\beta}\left(l^2+2\cancel l M_N + M_N^2\right)_{\gamma\delta}}{(l^2-M_N^2)^3}\,.
\end{align}
To obtain an estimate, we assume hierarchical masses and take the limit of $M_N\rightarrow \infty$ for the heaviest mass. Similarly, we neglect the lightest mass and set it to zero. Note that this is justified since no power-like UV or IR divergences occur in those limits. Consequently, only the heaviest and the second-to-heaviest mass contribute, and we obtain
\begin{equation}
    \mathcal{L}^{(6)}_{\cancel{L}} = \frac{1}{16\pi^2} \frac{y_1 y_2 y^2 \delta M_N}{\text{max}(M_h,M_\phi,M_N)^2 \text{max}_2(M_h,M_\phi,M_N)} f_\text{box}(M_\phi, M_N, M_h)^2 \mathcal{O}^{(6)}_{\cancel{L}} \,,
\end{equation}
where $\text{max}_2(M_h,M_\phi,M_N)$ stands for the second largest mass and the loop factor is again of order unity. With the typical mass hierarchy $M_\phi\ll M_h < M_N$ and the decay width of the inverse seesaw case
\begin{equation}
    \label{eq:app_tau_invS}
    \Gamma_{\chi_2\rightarrow \bar \chi_1 \nu \nu}^{-1}\equiv \tau =\frac{1280 \pi^3}{y_1^2 y_2^2} \delta^2\frac{M_\phi^4 M_N^2}{m_\nu^2\left( \epsilon M\right)^5}\,,
\end{equation} 
we reproduce the result in Eq.~\eqref{eq:Lvioldecays_pD}.
Another $\Delta L=2$ process involving a coupling of $\chi_1$ and $\chi_2$ to neutrinos via Higgs-VEV insertions, and a $W$-exchange to produce an $e^\pm$-pair is shown in Fig.~\ref{fig:W_exchange}. It can be estimated similarly to before, except that an additional $v^2$ appears as well as a factor of $g^2/M_W^2\sim 1/v^2$ due to the weak charged current. Both additions cancel each other and thus we obtain a contribution of comparable size as from the diagrams in Fig.~\ref{fig:box_diagram_L_violation}.

\bibliographystyle{JHEP}
\bibliography{bibliography}

\providecommand{\href}[2]{#2}\begingroup\raggedright\begin{thebibliography}{10}

\bibitem{Cirelli:2025rky}
M.~Cirelli, A.~Kar and H.~Shaikh, \emph{{Indirect searches for realistic sub-GeV Dark Matter models}},  \href{https://arxiv.org/abs/2508.03819}{{\ttfamily 2508.03819}}.

\bibitem{Slatyer:2017sev}
T.R.~Slatyer, \emph{{Indirect Detection of Dark Matter}},  in \emph{{Theoretical Advanced Study Institute in Elementary Particle Physics}: {Anticipating the Next Discoveries in Particle Physics}}, pp.~297--353, 2018, \href{https://doi.org/10.1142/9789813233348_0005}{DOI} [\href{https://arxiv.org/abs/1710.05137}{{\ttfamily 1710.05137}}].

\bibitem{Ibarra:2013cra}
A.~Ibarra, D.~Tran and C.~Weniger, \emph{{Indirect Searches for Decaying Dark Matter}}, \href{https://doi.org/10.1142/S0217751X13300408}{\emph{Int. J. Mod. Phys. A} {\bfseries 28} (2013) 1330040} [\href{https://arxiv.org/abs/1307.6434}{{\ttfamily 1307.6434}}].

\bibitem{Audren:2014bca}
B.~Audren, J.~Lesgourgues, G.~Mangano, P.D.~Serpico and T.~Tram, \emph{{Strongest model-independent bound on the lifetime of Dark Matter}}, \href{https://doi.org/10.1088/1475-7516/2014/12/028}{\emph{JCAP} {\bfseries 12} (2014) 028} [\href{https://arxiv.org/abs/1407.2418}{{\ttfamily 1407.2418}}].

\bibitem{Enqvist:2015ara}
K.~Enqvist, S.~Nadathur, T.~Sekiguchi and T.~Takahashi, \emph{{Decaying dark matter and the tension in $\sigma_8$}}, \href{https://doi.org/10.1088/1475-7516/2015/09/067}{\emph{JCAP} {\bfseries 09} (2015) 067} [\href{https://arxiv.org/abs/1505.05511}{{\ttfamily 1505.05511}}].

\bibitem{Berezhiani:2015yta}
Z.~Berezhiani, A.D.~Dolgov and I.I.~Tkachev, \emph{{Reconciling Planck results with low redshift astronomical measurements}}, \href{https://doi.org/10.1103/PhysRevD.92.061303}{\emph{Phys. Rev. D} {\bfseries 92} (2015) 061303} [\href{https://arxiv.org/abs/1505.03644}{{\ttfamily 1505.03644}}].

\bibitem{Poulin:2016nat}
V.~Poulin, P.D.~Serpico and J.~Lesgourgues, \emph{{A fresh look at linear cosmological constraints on a decaying dark matter component}}, \href{https://doi.org/10.1088/1475-7516/2016/08/036}{\emph{JCAP} {\bfseries 08} (2016) 036} [\href{https://arxiv.org/abs/1606.02073}{{\ttfamily 1606.02073}}].

\bibitem{Enqvist:2019tsa}
K.~Enqvist, S.~Nadathur, T.~Sekiguchi and T.~Takahashi, \emph{{Constraints on decaying dark matter from weak lensing and cluster counts}}, \href{https://doi.org/10.1088/1475-7516/2020/04/015}{\emph{JCAP} {\bfseries 04} (2020) 015} [\href{https://arxiv.org/abs/1906.09112}{{\ttfamily 1906.09112}}].

\bibitem{Bringmann:2018jpr}
T.~Bringmann, F.~Kahlhoefer, K.~Schmidt-Hoberg and P.~Walia, \emph{{Converting nonrelativistic dark matter to radiation}}, \href{https://doi.org/10.1103/PhysRevD.98.023543}{\emph{Phys. Rev. D} {\bfseries 98} (2018) 023543} [\href{https://arxiv.org/abs/1803.03644}{{\ttfamily 1803.03644}}].

\bibitem{Pandey:2019plg}
K.L.~Pandey, T.~Karwal and S.~Das, \emph{{Alleviating the $H_0$ and $\sigma_8$ anomalies with a decaying dark matter model}}, \href{https://doi.org/10.1088/1475-7516/2020/07/026}{\emph{JCAP} {\bfseries 07} (2020) 026} [\href{https://arxiv.org/abs/1902.10636}{{\ttfamily 1902.10636}}].

\bibitem{DES:2020mpv}
{\scshape DES} collaboration, \emph{{Constraints on dark matter to dark radiation conversion in the late universe with DES-Y1 and external data}}, \href{https://doi.org/10.1103/PhysRevD.103.123528}{\emph{Phys. Rev. D} {\bfseries 103} (2021) 123528} [\href{https://arxiv.org/abs/2011.04606}{{\ttfamily 2011.04606}}].

\bibitem{Nygaard:2020sow}
A.~Nygaard, T.~Tram and S.~Hannestad, \emph{{Updated constraints on decaying cold dark matter}}, \href{https://doi.org/10.1088/1475-7516/2021/05/017}{\emph{JCAP} {\bfseries 05} (2021) 017} [\href{https://arxiv.org/abs/2011.01632}{{\ttfamily 2011.01632}}].

\bibitem{Holm:2022kkd}
E.B.~Holm, L.~Herold, S.~Hannestad, A.~Nygaard and T.~Tram, \emph{{Decaying dark matter with profile likelihoods}}, \href{https://doi.org/10.1103/PhysRevD.107.L021303}{\emph{Phys. Rev. D} {\bfseries 107} (2023) L021303} [\href{https://arxiv.org/abs/2211.01935}{{\ttfamily 2211.01935}}].

\bibitem{Alvi:2022aam}
S.~Alvi, T.~Brinckmann, M.~Gerbino, M.~Lattanzi and L.~Pagano, \emph{{Do you smell something decaying? Updated linear constraints on decaying dark matter scenarios}}, \href{https://doi.org/10.1088/1475-7516/2022/11/015}{\emph{JCAP} {\bfseries 11} (2022) 015} [\href{https://arxiv.org/abs/2205.05636}{{\ttfamily 2205.05636}}].

\bibitem{Nygaard:2023gel}
A.~Nygaard, E.B.~Holm, T.~Tram and S.~Hannestad, \emph{{Decaying Dark Matter and the Hubble Tension}},  \href{https://arxiv.org/abs/2307.00418}{{\ttfamily 2307.00418}}.

\bibitem{Aoyama:2011ba}
S.~Aoyama, K.~Ichiki, D.~Nitta and N.~Sugiyama, \emph{{Formulation and constraints on decaying dark matter with finite mass daughter particles}}, \href{https://doi.org/10.1088/1475-7516/2011/09/025}{\emph{JCAP} {\bfseries 09} (2011) 025} [\href{https://arxiv.org/abs/1106.1984}{{\ttfamily 1106.1984}}].

\bibitem{Wang:2012eka}
M.-Y.~Wang and A.R.~Zentner, \emph{{Effects of Unstable Dark Matter on Large-Scale Structure and Constraints from Future Surveys}}, \href{https://doi.org/10.1103/PhysRevD.85.043514}{\emph{Phys. Rev. D} {\bfseries 85} (2012) 043514} [\href{https://arxiv.org/abs/1201.2426}{{\ttfamily 1201.2426}}].

\bibitem{Wang:2014ina}
M.-Y.~Wang, A.H.G.~Peter, L.E.~Strigari, A.R.~Zentner, B.~Arant, S.~Garrison-Kimmel et~al., \emph{{Cosmological simulations of decaying dark matter: implications for small-scale structure of dark matter haloes}}, \href{https://doi.org/10.1093/mnras/stu1747}{\emph{Mon. Not. Roy. Astron. Soc.} {\bfseries 445} (2014) 614} [\href{https://arxiv.org/abs/1406.0527}{{\ttfamily 1406.0527}}].

\bibitem{Blackadder:2014wpa}
G.~Blackadder and S.M.~Koushiappas, \emph{{Dark matter with two- and many-body decays and supernovae type Ia}}, \href{https://doi.org/10.1103/PhysRevD.90.103527}{\emph{Phys. Rev. D} {\bfseries 90} (2014) 103527} [\href{https://arxiv.org/abs/1410.0683}{{\ttfamily 1410.0683}}].

\bibitem{Aoyama:2014tga}
S.~Aoyama, T.~Sekiguchi, K.~Ichiki and N.~Sugiyama, \emph{{Evolution of perturbations and cosmological constraints in decaying dark matter models with arbitrary decay mass products}}, \href{https://doi.org/10.1088/1475-7516/2014/07/021}{\emph{JCAP} {\bfseries 07} (2014) 021} [\href{https://arxiv.org/abs/1402.2972}{{\ttfamily 1402.2972}}].

\bibitem{Blackadder:2015uta}
G.~Blackadder and S.M.~Koushiappas, \emph{{Cosmological constraints to dark matter with two- and many-body decays}}, \href{https://doi.org/10.1103/PhysRevD.93.023510}{\emph{Phys. Rev. D} {\bfseries 93} (2016) 023510} [\href{https://arxiv.org/abs/1510.06026}{{\ttfamily 1510.06026}}].

\bibitem{Vattis:2019efj}
K.~Vattis, S.M.~Koushiappas and A.~Loeb, \emph{{Dark matter decaying in the late Universe can relieve the H0 tension}}, \href{https://doi.org/10.1103/PhysRevD.99.121302}{\emph{Phys. Rev. D} {\bfseries 99} (2019) 121302} [\href{https://arxiv.org/abs/1903.06220}{{\ttfamily 1903.06220}}].

\bibitem{PhysRevD.103.043014}
S.J.~Clark, K.~Vattis and S.M.~Koushiappas, \emph{Cosmological constraints on late-universe decaying dark matter as a solution to the ${H}_{0}$ tension}, \href{https://doi.org/10.1103/PhysRevD.103.043014}{\emph{Phys. Rev. D} {\bfseries 103} (2021) 043014}.

\bibitem{Haridasu:2020xaa}
B.S.~Haridasu and M.~Viel, \emph{{Late-time decaying dark matter: constraints and implications for the $H_0$-tension}}, \href{https://doi.org/10.1093/mnras/staa1991}{\emph{Mon. Not. Roy. Astron. Soc.} {\bfseries 497} (2020) 1757} [\href{https://arxiv.org/abs/2004.07709}{{\ttfamily 2004.07709}}].

\bibitem{Abellan:2020pmw}
G.F.~Abell\'an, R.~Murgia, V.~Poulin and J.~Lavalle, \emph{{Implications of the $S_8$ tension for decaying dark matter with warm decay products}}, \href{https://doi.org/10.1103/PhysRevD.105.063525}{\emph{Phys. Rev. D} {\bfseries 105} (2022) 063525} [\href{https://arxiv.org/abs/2008.09615}{{\ttfamily 2008.09615}}].

\bibitem{FrancoAbellan:2020xnr}
G.~Franco~Abell{\'a}n, R.~Murgia, V.~Poulin and J.~Lavalle, \emph{{Implications of the $S_8$ tension for decaying dark matter with warm decay products}}, \href{https://doi.org/10.1103/PhysRevD.105.063525}{\emph{Phys. Rev. D} {\bfseries 105} (2022) 063525} [\href{https://arxiv.org/abs/2008.09615}{{\ttfamily 2008.09615}}].

\bibitem{FrancoAbellan:2021sxk}
G.~Franco~Abell\'an, R.~Murgia and V.~Poulin, \emph{{Linear cosmological constraints on two-body decaying dark matter scenarios and the S8 tension}}, \href{https://doi.org/10.1103/PhysRevD.104.123533}{\emph{Phys. Rev. D} {\bfseries 104} (2021) 123533} [\href{https://arxiv.org/abs/2102.12498}{{\ttfamily 2102.12498}}].

\bibitem{Simon:2022ftd}
T.~Simon, G.~Franco~Abell\'an, P.~Du, V.~Poulin and Y.~Tsai, \emph{{Constraining decaying dark matter with BOSS data and the effective field theory of large-scale structures}}, \href{https://doi.org/10.1103/PhysRevD.106.023516}{\emph{Phys. Rev. D} {\bfseries 106} (2022) 023516} [\href{https://arxiv.org/abs/2203.07440}{{\ttfamily 2203.07440}}].

\bibitem{Fuss:2022zyt}
L.~Fu\ss{} and M.~Garny, \emph{{Decaying Dark Matter and Lyman-\ensuremath{\alpha} forest constraints}}, \href{https://doi.org/10.1088/1475-7516/2023/10/020}{\emph{JCAP} {\bfseries 10} (2023) 020} [\href{https://arxiv.org/abs/2210.06117}{{\ttfamily 2210.06117}}].

\bibitem{Bucko:2023eix}
J.~Bucko, S.K.~Giri, F.H.~Peters and A.~Schneider, \emph{{Probing the two-body decaying dark matter scenario with weak lensing and the cosmic microwave background}},  \href{https://arxiv.org/abs/2307.03222}{{\ttfamily 2307.03222}}.

\bibitem{Montandon:2025xpd}
T.~Montandon, E.M.~Teixeira, A.~Poudou and V.~Poulin, \emph{{A frequentist view on the two-body decaying dark matter model}},  \href{https://arxiv.org/abs/2505.20193}{{\ttfamily 2505.20193}}.

\bibitem{Peter:2010sz}
A.H.G.~Peter and A.J.~Benson, \emph{{Dark-matter decays and Milky Way satellite galaxies}}, \href{https://doi.org/10.1103/PhysRevD.82.123521}{\emph{Phys. Rev. D} {\bfseries 82} (2010) 123521} [\href{https://arxiv.org/abs/1009.1912}{{\ttfamily 1009.1912}}].

\bibitem{DES:2022doi}
{\scshape DES} collaboration, \emph{{Milky Way Satellite Census. IV. Constraints on Decaying Dark Matter from Observations of Milky Way Satellite Galaxies}}, \href{https://doi.org/10.3847/1538-4357/ac6e65}{\emph{Astrophys. J.} {\bfseries 932} (2022) 128} [\href{https://arxiv.org/abs/2201.11740}{{\ttfamily 2201.11740}}].

\bibitem{Lester:2025hqt}
E.~Lester and K.~Bolejko, \emph{{Constraining decaying dark matter models with gravitational lensing and cosmic voids}},  \href{https://arxiv.org/abs/2507.08275}{{\ttfamily 2507.08275}}.

\bibitem{Davari:2022uwd}
Z.~Davari and N.~Khosravi, \emph{{Can decaying dark matter scenarios alleviate both H0 and \ensuremath{\sigma}8 tensions?}}, \href{https://doi.org/10.1093/mnras/stac2306}{\emph{Mon. Not. Roy. Astron. Soc.} {\bfseries 516} (2022) 4373} [\href{https://arxiv.org/abs/2203.09439}{{\ttfamily 2203.09439}}].

\bibitem{Zhou:2025ikl}
Q.~Zhou, Z.~Xu and S.~Zheng, \emph{{Interpreting Hubble tension with a cascade decaying dark matter sector}},  \href{https://arxiv.org/abs/2507.08687}{{\ttfamily 2507.08687}}.

\bibitem{Lynch:2025ine}
G.P.~Lynch and L.~Knox, \emph{{What's the matter with $\Sigma m_{\nu}$?}},  \href{https://arxiv.org/abs/2503.14470}{{\ttfamily 2503.14470}}.

\bibitem{Giare:2025ath}
W.~Giar{\`e}, O.~Mena, E.~Specogna and E.~Di~Valentino, \emph{{Neutrino mass tension or suppressed growth rate of matter perturbations?}},  \href{https://arxiv.org/abs/2507.01848}{{\ttfamily 2507.01848}}.

\bibitem{Fuss:2024dam}
L.~Fu\ss{}, M.~Garny and A.~Ibarra, \emph{{Minimal decaying dark matter: from cosmological tensions to neutrino signatures}},  \href{https://arxiv.org/abs/2403.15543}{{\ttfamily 2403.15543}}.

\bibitem{ColomerMolla:2023ppf}
{\scshape JUNO} collaboration, \emph{{Physics potential with astrophysical neutrinos in JUNO}}, \href{https://doi.org/10.22323/1.444.1192}{\emph{PoS} {\bfseries ICRC2023} (2023) 1192}.

\bibitem{Nadler:2025yni}
E.O.~Nadler and A.J.~Benson, \emph{{Semianalytic model for decaying dark matter halos}}, \href{https://doi.org/10.1103/PhysRevD.111.103522}{\emph{Phys. Rev. D} {\bfseries 111} (2025) 103522} [\href{https://arxiv.org/abs/2501.12636}{{\ttfamily 2501.12636}}].

\bibitem{Bell:2010fk}
N.F.~Bell, A.J.~Galea and K.~Petraki, \emph{{Lifetime Constraints for Late Dark Matter Decay}}, \href{https://doi.org/10.1103/PhysRevD.82.023514}{\emph{Phys. Rev. D} {\bfseries 82} (2010) 023514} [\href{https://arxiv.org/abs/1004.1008}{{\ttfamily 1004.1008}}].

\bibitem{Bell:2010qt}
N.F.~Bell, A.J.~Galea and R.R.~Volkas, \emph{{A Model For Late Dark Matter Decay}}, \href{https://doi.org/10.1103/PhysRevD.83.063504}{\emph{Phys. Rev. D} {\bfseries 83} (2011) 063504} [\href{https://arxiv.org/abs/1012.0067}{{\ttfamily 1012.0067}}].

\bibitem{Hamaguchi:2017ihw}
K.~Hamaguchi, K.~Nakayama and Y.~Tang, \emph{{Gravitino/Axino as Decaying Dark Matter and Cosmological Tensions}}, \href{https://doi.org/10.1016/j.physletb.2017.06.071}{\emph{Phys. Lett. B} {\bfseries 772} (2017) 415} [\href{https://arxiv.org/abs/1705.04521}{{\ttfamily 1705.04521}}].

\bibitem{Bae:2018mgq}
K.J.~Bae, A.~Kamada and H.J.~Kim, \emph{{Decaying axinolike dark matter: Discriminative solution to small-scale issues}}, \href{https://doi.org/10.1103/PhysRevD.99.023511}{\emph{Phys. Rev. D} {\bfseries 99} (2019) 023511} [\href{https://arxiv.org/abs/1806.08569}{{\ttfamily 1806.08569}}].

\bibitem{Choi:2021uhy}
G.~Choi and T.T.~Yanagida, \emph{{Gravitino cosmology helped by a right handed (s)neutrino}}, \href{https://doi.org/10.1016/j.physletb.2022.136954}{\emph{Phys. Lett. B} {\bfseries 827} (2022) 136954} [\href{https://arxiv.org/abs/2104.02958}{{\ttfamily 2104.02958}}].

\bibitem{Deshpande:2023gij}
M.~Deshpande, \emph{{A study of supersymmetric decaying dark matter models}}, Ph.D. thesis, Adelaide U., 2023.

\bibitem{Obied:2023clp}
G.~Obied, C.~Dvorkin, E.~Gonzalo and C.~Vafa, \emph{{Dark Dimension and Decaying Dark Matter Gravitons}},  \href{https://arxiv.org/abs/2311.05318}{{\ttfamily 2311.05318}}.

\bibitem{Cheek:2022yof}
A.~Cheek, J.K.~Osi\'nski, L.~Roszkowski and S.~Trojanowski, \emph{{Dark matter production through a non-thermal flavon portal}}, \href{https://doi.org/10.1007/JHEP03(2023)149}{\emph{JHEP} {\bfseries 03} (2023) 149} [\href{https://arxiv.org/abs/2211.02057}{{\ttfamily 2211.02057}}].

\bibitem{Cheek:2024fyc}
A.~Cheek, Y.-C.~Qiu and L.~Tan, \emph{{Gemini dark matter}}, \href{https://doi.org/10.1103/PhysRevD.110.075021}{\emph{Phys. Rev. D} {\bfseries 110} (2024) 075021} [\href{https://arxiv.org/abs/2407.01099}{{\ttfamily 2407.01099}}].

\bibitem{Cardenas:2024ojd}
K.M.~C{\'a}rdenas, G.~Mohlabeng and A.C.~Vincent, \emph{{Global fit to loopy dark matter and neutrino masses}}, \href{https://doi.org/10.1103/PhysRevD.111.055024}{\emph{Phys. Rev. D} {\bfseries 111} (2025) 055024} [\href{https://arxiv.org/abs/2411.03470}{{\ttfamily 2411.03470}}].

\bibitem{DelaTorreLuque:2023olp}
P.~De~la Torre~Luque, S.~Balaji and J.~Koechler, \emph{{Importance of cosmic ray propagation on sub-GeV dark matter constraints}},  \href{https://arxiv.org/abs/2311.04979}{{\ttfamily 2311.04979}}.

\bibitem{DelaTorreLuque:2023cef}
P.~De~la Torre~Luque, S.~Balaji and J.~Silk, \emph{{New 511 keV line data provides strongest sub-GeV dark matter constraints}},  \href{https://arxiv.org/abs/2312.04907}{{\ttfamily 2312.04907}}.

\bibitem{Jin:2013nta}
H.-B.~Jin, Y.-L.~Wu and Y.-F.~Zhou, \emph{{Implications of the first AMS-02 measurement for dark matter annihilation and decay}}, \href{https://doi.org/10.1088/1475-7516/2013/11/026}{\emph{JCAP} {\bfseries 11} (2013) 026} [\href{https://arxiv.org/abs/1304.1997}{{\ttfamily 1304.1997}}].

\bibitem{Mohapatra:1979ia}
R.N.~Mohapatra and G.~Senjanovic, \emph{{Neutrino Mass and Spontaneous Parity Nonconservation}}, \href{https://doi.org/10.1103/PhysRevLett.44.912}{\emph{Phys. Rev. Lett.} {\bfseries 44} (1980) 912}.

\bibitem{Minkowski:1977sc}
P.~Minkowski, \emph{{$\mu \to e\gamma$ at a Rate of One Out of $10^{9}$ Muon Decays?}}, \href{https://doi.org/10.1016/0370-2693(77)90435-X}{\emph{Phys. Lett. B} {\bfseries 67} (1977) 421}.

\bibitem{Yanagida:1979as}
T.~Yanagida, \emph{{Horizontal gauge symmetry and masses of neutrinos}}, {\emph{Conf. Proc. C} {\bfseries 7902131} (1979) 95}.

\bibitem{Gell-Mann:1979vob}
M.~Gell-Mann, P.~Ramond and R.~Slansky, \emph{{Complex Spinors and Unified Theories}}, {\emph{Conf. Proc. C} {\bfseries 790927} (1979) 315} [\href{https://arxiv.org/abs/1306.4669}{{\ttfamily 1306.4669}}].

\bibitem{Wyler:1982dd}
D.~Wyler and L.~Wolfenstein, \emph{{Massless Neutrinos in Left-Right Symmetric Models}}, \href{https://doi.org/10.1016/0550-3213(83)90482-0}{\emph{Nucl. Phys. B} {\bfseries 218} (1983) 205}.

\bibitem{Mohapatra:1986aw}
R.N.~Mohapatra, \emph{{Mechanism for Understanding Small Neutrino Mass in Superstring Theories}}, \href{https://doi.org/10.1103/PhysRevLett.56.561}{\emph{Phys. Rev. Lett.} {\bfseries 56} (1986) 561}.

\bibitem{Mohapatra:1986bd}
R.N.~Mohapatra and J.W.F.~Valle, \emph{{Neutrino Mass and Baryon Number Nonconservation in Superstring Models}}, \href{https://doi.org/10.1103/PhysRevD.34.1642}{\emph{Phys. Rev. D} {\bfseries 34} (1986) 1642}.

\bibitem{Bernabeu:1987gr}
J.~Bernabeu, A.~Santamaria, J.~Vidal, A.~Mendez and J.W.F.~Valle, \emph{{Lepton Flavor Nonconservation at High-Energies in a Superstring Inspired Standard Model}}, \href{https://doi.org/10.1016/0370-2693(87)91100-2}{\emph{Phys. Lett. B} {\bfseries 187} (1987) 303}.

\bibitem{Akita:2022lit}
K.~Akita, G.~Lambiase, M.~Niibo and M.~Yamaguchi, \emph{{Neutrino lines from MeV dark matter annihilation and decay in JUNO}}, \href{https://doi.org/10.1088/1475-7516/2022/10/097}{\emph{JCAP} {\bfseries 10} (2022) 097} [\href{https://arxiv.org/abs/2206.06755}{{\ttfamily 2206.06755}}].

\bibitem{KiDS:2020suj}
{\scshape KiDS} collaboration, \emph{{KiDS-1000 Cosmology: Cosmic shear constraints and comparison between two point statistics}}, \href{https://doi.org/10.1051/0004-6361/202039070}{\emph{Astron. Astrophys.} {\bfseries 645} (2021) A104} [\href{https://arxiv.org/abs/2007.15633}{{\ttfamily 2007.15633}}].

\bibitem{Wright:2025xka}
A.H.~Wright et~al., \emph{{KiDS-Legacy: Cosmological constraints from cosmic shear with the complete Kilo-Degree Survey}},  \href{https://arxiv.org/abs/2503.19441}{{\ttfamily 2503.19441}}.

\bibitem{Planck:2018vyg}
{\scshape Planck} collaboration, \emph{{Planck 2018 results. VI. Cosmological parameters}}, \href{https://doi.org/10.1051/0004-6361/201833910}{\emph{Astron. Astrophys.} {\bfseries 641} (2020) A6} [\href{https://arxiv.org/abs/1807.06209}{{\ttfamily 1807.06209}}].

\bibitem{Abdalla:2022yfr}
E.~Abdalla et~al., \emph{{Cosmology intertwined: A review of the particle physics, astrophysics, and cosmology associated with the cosmological tensions and anomalies}}, \href{https://doi.org/10.1016/j.jheap.2022.04.002}{\emph{JHEAp} {\bfseries 34} (2022) 49} [\href{https://arxiv.org/abs/2203.06142}{{\ttfamily 2203.06142}}].

\bibitem{Doux:2025vru}
C.~Doux and T.~Karwal, \emph{{Going beyond $S_8$: fast inference of the matter power spectrum from weak-lensing surveys}},  \href{https://arxiv.org/abs/2506.16434}{{\ttfamily 2506.16434}}.

\bibitem{DES:2021wwk}
{\scshape DES} collaboration, \emph{{Dark Energy Survey Year 3 results: Cosmological constraints from galaxy clustering and weak lensing}}, \href{https://doi.org/10.1103/PhysRevD.105.023520}{\emph{Phys. Rev. D} {\bfseries 105} (2022) 023520} [\href{https://arxiv.org/abs/2105.13549}{{\ttfamily 2105.13549}}].

\bibitem{Sugiyama:2023fzm}
S.~Sugiyama et~al., \emph{{Hyper Suprime-Cam Year 3 results: Cosmology from galaxy clustering and weak lensing with HSC and SDSS using the minimal bias model}}, \href{https://doi.org/10.1103/PhysRevD.108.123521}{\emph{Phys. Rev. D} {\bfseries 108} (2023) 123521} [\href{https://arxiv.org/abs/2304.00705}{{\ttfamily 2304.00705}}].

\bibitem{2024AstHe.117..304S}
S.~{Sugiyama}, M.~{Takada} and H.~{Miyatake}, \emph{{Weak Lensing Cosmology with Subaru HSC Data}}, {\emph{Astronomical Herald} {\bfseries 117} (2024) 304}.

\bibitem{SPT:2018njh}
{\scshape SPT} collaboration, \emph{{Cluster Cosmology Constraints from the 2500 deg$^2$ SPT-SZ Survey: Inclusion of Weak Gravitational Lensing Data from Magellan and the Hubble Space Telescope}}, \href{https://doi.org/10.3847/1538-4357/ab1f10}{\emph{Astrophys. J.} {\bfseries 878} (2019) 55} [\href{https://arxiv.org/abs/1812.01679}{{\ttfamily 1812.01679}}].

\bibitem{Ivanov:2019pdj}
M.M.~Ivanov, M.~Simonovi{\'c} and M.~Zaldarriaga, \emph{{Cosmological Parameters from the BOSS Galaxy Power Spectrum}}, \href{https://doi.org/10.1088/1475-7516/2020/05/042}{\emph{JCAP} {\bfseries 05} (2020) 042} [\href{https://arxiv.org/abs/1909.05277}{{\ttfamily 1909.05277}}].

\bibitem{Fernandez-Martinez:2023phj}
E.~Fern\'andez-Mart\'\i{}nez, M.~Gonz\'alez-L\'opez, J.~Hern\'andez-Garc\'\i{}a, M.~Hostert and J.~L\'opez-Pav\'on, \emph{{Effective portals to heavy neutral leptons}}, \href{https://doi.org/10.1007/JHEP09(2023)001}{\emph{JHEP} {\bfseries 09} (2023) 001} [\href{https://arxiv.org/abs/2304.06772}{{\ttfamily 2304.06772}}].

\bibitem{Bechtol:2022koa}
K.~Bechtol et~al., \emph{{Snowmass2021 Cosmic Frontier White Paper: Dark Matter Physics from Halo Measurements}},  in \emph{{2022 Snowmass Summer Study}}, 3, 2022 [\href{https://arxiv.org/abs/2203.07354}{{\ttfamily 2203.07354}}].

\bibitem{Bondarenko:2018}
K.~Bondarenko, A.~Boyarsky, D.~Gorbunov and O.~Ruchayskiy, \emph{Phenomenology of gev-scale heavy neutral leptons}, \href{https://doi.org/10.1007/JHEP11(2018)032}{\emph{Journal of High Energy Physics} {\bfseries 2018} (2018) }.

\bibitem{Vincent:2014rja}
A.C.~Vincent, E.F.~Martinez, P.~Hern\'andez, M.~Lattanzi and O.~Mena, \emph{{Revisiting cosmological bounds on sterile neutrinos}}, \href{https://doi.org/10.1088/1475-7516/2015/04/006}{\emph{JCAP} {\bfseries 04} (2015) 006} [\href{https://arxiv.org/abs/1408.1956}{{\ttfamily 1408.1956}}].

\bibitem{Dev:2025pru}
P.S.B.~Dev, Q.-f.~Wu and X.-J.~Xu, \emph{{No Hiding in the Dark: Cosmological Bounds on Heavy Neutral Leptons with Dark Decay Channels}},  \href{https://arxiv.org/abs/2507.12270}{{\ttfamily 2507.12270}}.

\bibitem{Boyarsky:2020dzc}
A.~Boyarsky, M.~Ovchynnikov, O.~Ruchayskiy and V.~Syvolap, \emph{{Improved big bang nucleosynthesis constraints on heavy neutral leptons}}, \href{https://doi.org/10.1103/PhysRevD.104.023517}{\emph{Phys. Rev. D} {\bfseries 104} (2021) 023517} [\href{https://arxiv.org/abs/2008.00749}{{\ttfamily 2008.00749}}].

\bibitem{Ghisoiu:2014ena}
I.~Ghisoiu and M.~Laine, \emph{{Right-handed neutrino production rate at T {\ensuremath{>}} 160 GeV}}, \href{https://doi.org/10.1088/1475-7516/2014/12/032}{\emph{JCAP} {\bfseries 12} (2014) 032} [\href{https://arxiv.org/abs/1411.1765}{{\ttfamily 1411.1765}}].

\bibitem{Ghiglieri:2016xye}
J.~Ghiglieri and M.~Laine, \emph{{Neutrino dynamics below the electroweak crossover}}, \href{https://doi.org/10.1088/1475-7516/2016/07/015}{\emph{JCAP} {\bfseries 07} (2016) 015} [\href{https://arxiv.org/abs/1605.07720}{{\ttfamily 1605.07720}}].

\bibitem{Alonso-Alvarez:2022uxp}
G.~Alonso-{\'A}lvarez and J.M.~Cline, \emph{{Sterile neutrino production at small mixing in the early universe}}, \href{https://doi.org/10.1016/j.physletb.2022.137278}{\emph{Phys. Lett. B} {\bfseries 833} (2022) 137278} [\href{https://arxiv.org/abs/2204.04224}{{\ttfamily 2204.04224}}].

\bibitem{Dodelson:1993je}
S.~Dodelson and L.M.~Widrow, \emph{{Sterile-neutrinos as dark matter}}, \href{https://doi.org/10.1103/PhysRevLett.72.17}{\emph{Phys. Rev. Lett.} {\bfseries 72} (1994) 17} [\href{https://arxiv.org/abs/hep-ph/9303287}{{\ttfamily hep-ph/9303287}}].

\bibitem{Dolgov:2000ew}
A.D.~Dolgov and S.H.~Hansen, \emph{{Massive sterile neutrinos as warm dark matter}}, \href{https://doi.org/10.1016/S0927-6505(01)00115-3}{\emph{Astropart. Phys.} {\bfseries 16} (2002) 339} [\href{https://arxiv.org/abs/hep-ph/0009083}{{\ttfamily hep-ph/0009083}}].

\bibitem{Ibarra:2013zia}
A.~Ibarra, A.S.~Lamperstorfer and J.~Silk, \emph{{Dark matter annihilations and decays after the AMS-02 positron measurements}}, \href{https://doi.org/10.1103/PhysRevD.89.063539}{\emph{Phys. Rev. D} {\bfseries 89} (2014) 063539} [\href{https://arxiv.org/abs/1309.2570}{{\ttfamily 1309.2570}}].

\bibitem{Garny:2010eg}
M.~Garny, A.~Ibarra, D.~Tran and C.~Weniger, \emph{{Gamma-Ray Lines from Radiative Dark Matter Decay}}, \href{https://doi.org/10.1088/1475-7516/2011/01/032}{\emph{JCAP} {\bfseries 01} (2011) 032} [\href{https://arxiv.org/abs/1011.3786}{{\ttfamily 1011.3786}}].

\bibitem{Tomsick:2019wvo}
J.A.~Tomsick et~al., \emph{{The Compton Spectrometer and Imager}},  \href{https://arxiv.org/abs/1908.04334}{{\ttfamily 1908.04334}}.

\bibitem{Beechert:2022phz}
J.~Beechert et~al., \emph{{Calibrations of the Compton Spectrometer and Imager}}, \href{https://doi.org/10.1016/j.nima.2022.166510}{\emph{Nucl. Instrum. Meth. A} {\bfseries 1031} (2022) 166510} [\href{https://arxiv.org/abs/2203.00695}{{\ttfamily 2203.00695}}].

\end{thebibliography}\endgroup

\end{document}